\documentclass[epsf,prx,twocolumn,showpacs,nofootinbib,longbibliography]{revtex4-1}

\usepackage[pdftex]{graphicx}
\usepackage{dcolumn}
\usepackage{bm}
\usepackage{epsfig}
\usepackage{latexsym}
\usepackage{amsmath}
\usepackage{amsfonts}
\usepackage{amssymb}
\usepackage{float}
\usepackage{caption}
\usepackage{graphicx}
\graphicspath{{images/}}
\usepackage{subcaption}
\usepackage{color}
\usepackage{framed}
\usepackage{array}
\newcolumntype{P}[1]{>{\centering\arraybackslash}p{#1}}

\usepackage{makecell}

\begin{document}

\title{Localization in fractonic random circuits}
\author{Shriya Pai, Michael Pretko, and Rahul M. Nandkishore}
\affiliation{Department of Physics and Center for Theory of Quantum Matter, University of Colorado, Boulder, CO 80309, USA}
\date{\today}

\begin{abstract} 
We study the spreading of initially-local operators under unitary time evolution in a one-dimensional random quantum circuit model which is constrained to conserve a $U(1)$ charge and also the dipole moment of this charge. These constraints are motivated by the quantum dynamics of fracton phases. We discover that charge remains localized at its initial position, providing a crisp example of a non-ergodic dynamical phase of random circuit dynamics. This localization can be understood as a consequence of the return properties of low dimensional random walks, through a  mechanism reminiscent of weak localization, but insensitive to dephasing. The charge dynamics is well-described by a system of coupled hydrodynamic equations, which makes several non-trivial predictions which are all in good agreement with numerics in one dimension. Importantly, these equations also predict localization in {\it two}-dimensional fractonic random circuits. We further find that the immobile fractonic charge emits non-conserved operators, whose spreading is governed by exponents distinct from those observed in non-fractonic circuits. These non-standard exponents are also explained by our coupled hydrodynamic equations. Where entanglement properties are concerned, we find that fractonic operators exhibit a short time linear growth of observable entanglement with saturation to an {\it area} law, as well as a subthermal volume law for operator entanglement. The entanglement spectrum is found to follow semi-Poisson statistics, similar to eigenstates of many-body localized systems. The non-ergodic phenomenology is found to persist to initial conditions containing non-zero density of dipolar or fractonic charge, including states near the sector of maximal charge. Our work implies that low-dimensional fracton systems should preserve forever a memory of their initial conditions in local observables under noisy quantum dynamics, thereby constituting ideal memories. It also implies that one- and two-dimensional fracton systems should realize true many-body localization (MBL) under Hamiltonian dynamics, even in the absence of disorder, with the obstructions to MBL in translation invariant systems and in spatial dimensions greater than one being evaded by the nature of the mechanism responsible for localization. We also suggest a possible route to new non-ergodic phases in high dimensions. 
\end{abstract}
\maketitle

\normalsize

\tableofcontents

\section{Introduction}
\begin{figure*}[t]
 \centering
 \includegraphics[scale=0.5]{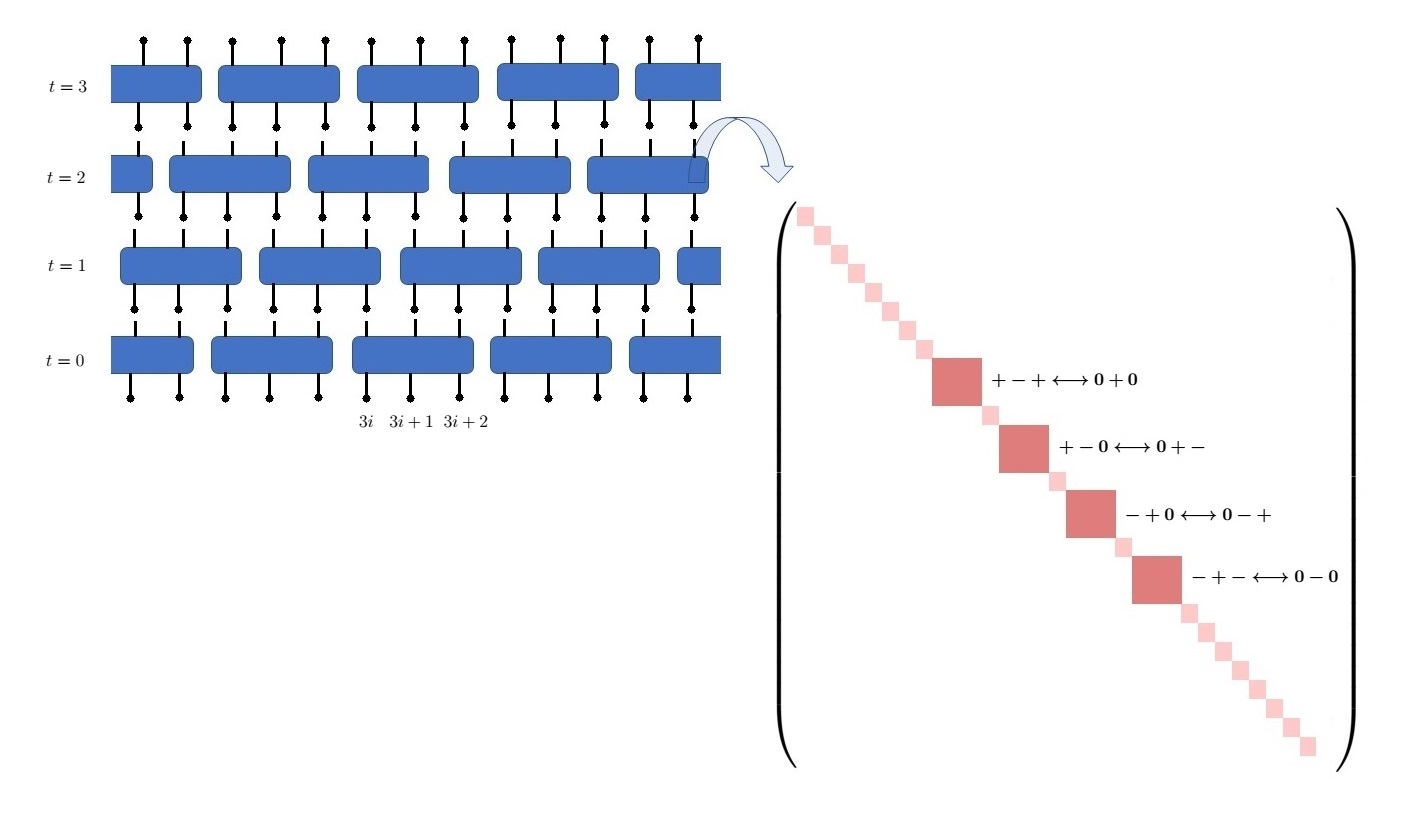}
\captionsetup{justification=centering,margin=1cm}
 \caption{Random unitary circuit: each site is a three-state qudit. Each gate (blue box) locally conserves
$S_{z}^{\textrm{total}}$, the total $z$ component of the three qudits it acts upon, and $\vec{P}_{\textrm{total}}$, the total dipole moment of the three qudits. It is a block-diagonal unitary of the form shown on the right, with each block (red boxes) of each gate independently Haar-random. The smaller $1 \times 1$ blocks do not flip the qudits, while the larger $2 \times 2$ blocks produce charge- and dipole-conserving qudit flips as indicated.}
 \label{fig:randomunitary}
 \end{figure*}
The quantum dynamics of interacting many-body systems is a great open frontier for theoretical physics. Important advances on this front over the past decade include the development of the theories of many-body localization (MBL) \cite{GMP, BAA, mblarcmp, AABS} and the eigenstate thermalization hypothesis (ETH) \cite{Deutsch, Rigol, Srednicki}, and advances in our understanding of the scrambling of information in quantum chaos \cite{SY, K, Maldacena}. While these advances have occurred in the context of closed quantum systems with time translation symmetry (either continuous or discrete), more recent advances have also shed light on our understanding of quantum dynamics subject to {\it noise} i.e. without time translation symmetry, and hence without the constraints imposed by energy conservation. These advances stem from the exploration of many-body dynamics in random unitary {\it circuits} \footnote{For a discussion of how random circuit dynamics can be obtained by applying random pulses to a many-body system, see \cite{banchi}.}, where the time evolution is generated by the application of random gates \cite{Nahum1, Nahum2, Nahum3, Nahum4, Keyserlingk1, Keyserlingk2, KhemaniVishHuse}.  (Similar work has also been done in the context of Floquet circuits, featuring periodic time evolution \cite{Prosen98, Prosen99, Prosen, amos1,amos2,HuseCirac}.)  These explorations have provided an unprecedentedly detailed understanding of the onset of many-body quantum chaos. However, all random circuit models that have been studied so far thermalize \footnote{Non-ergodic dynamics are possible if one restricts to more structured settings such as Floquet systems \cite{Prosen98, Prosen99, amos2} or Clifford circuits  \cite{GopalakrishnanClifford}. }, and it remains unclear whether noisy quantum dynamics can lead to robust {\it non-ergodic} phases analogous to MBL. 

 \begin{table*}[t]
  \centering
\begin{tabular}{ |P{3cm}||P{4cm}|P{6cm}|  }
 \hline
 Net charge & 3-qudit gates & 4-qudit gates\\
 \hline
 $+2$   &    & \makecell{$+\ 0\ 0\ + \leftrightarrow 0\ +\ +\ 0$                                                                                                                                                                                                                                                                                                                                                                                                                                                    \\$0\ +\ 0\ + \leftrightarrow +\ -\ +\ +$                                                                                                                                                                                                                                                                                                                                                                                                                                                  \\$+\ 0\ +\ 0 \leftrightarrow +\ +\ -\ +$                                                                                                                                                                                                                                                                                                                                                                                                                                                  } \\
 \hline
  $+1$  & $+ - + \leftrightarrow 0 + 0$ & \makecell{$0\ +\ 0\ 0 \leftrightarrow +\ -\ +\ 0 \leftrightarrow +\ 0\ -\ +$                                                                                                                                                                                                                                                                                                                                                                                                                                                  \\$0\ 0\ +\ 0 \leftrightarrow +\ -\ 0\ + \leftrightarrow 0\ +\ -\ +$                                                                                                                                                                                                                                                                                                                                                                                                                                                  \\
$+\ 0\ 0\ 0 \leftrightarrow 0\ +\ +\ -$                                                                                                                                                                                                                                                                                                                                                                                                                                                   \\$0\ 0\ 0\ + \leftrightarrow -\ +\ +\ 0$\\
$-\ +\ 0\ + \leftrightarrow 0\ -\ +\ +$\\
$+\ +\ -\ 0 \leftrightarrow +\ 0\ +\ -$                                                                                                                                                                                                                                                                                                                                                                                                                                                                                                                                                                                                                                                                                                                                                                                                                                                                                                                                                                                                                                                                                                                                                                                                                                                                                                                                                      }\\
\hline
 $0$ & $-\ +\ 0 \leftrightarrow 0\ -\ +$ & \makecell{$0\ 0\ 0\ 0 \leftrightarrow +\ -\ -\ + \leftrightarrow -\ +\ +\ -$\\
 $-\ +\ 0\ 0 \leftrightarrow 0\ -\ +\ 0 \leftrightarrow 0\ 0\ -\ +$\\
 $+\ -\ 0\ 0 \leftrightarrow 0\ +\ -\ 0 \leftrightarrow 0\ 0\ +\ -$\\
$+\ 0\ -\ 0 \leftrightarrow 0\ +\ 0\ -$                                                                                                                                                                                                                                                                                                                                                                                                                                                                                                                                                                                                                                                                                                                                                                                                                                                                                                                                                                                                                                                                                                                                                                                                                                                                                                                                                     \\
$+\ 0\ 0\ - \leftrightarrow +\ -\ +\ -$                                                                                                                                                                                                                                                                                                                                                                                                                                                                                                                                                                                                                                                                                                                                                                                                                                                                                                                                                                                                                                                                                                                                                                                                                                                                                                                                                     }\\
\hline

 \end{tabular}
\caption{Gates implementing flips used in constructing nontrivial blocks of 3- and 4-site unitaries. The remaining gates for net charge $0$ are to be obtained by $+ \leftrightarrow -$ in only its $2 \times 2$ blocks, and the remaining negative net charge gates are to be obtained by $+ \leftrightarrow -$ in all blocks of the corresponding positive charge gates. The blocks forming the 5-site unitary matrix are obtained in a similar fashion.}
  \label{qudit flip gates}
\end{table*}

Parallel to the developments in quantum dynamics, recent years have also witnessed an explosion of interest in {\it fracton} phases of quantum matter \cite{chamon, castelnovo, haah, fracton1, fracton2, sub, genem, han, sagar, slagle1, slagle2, MaSchmitz, balents, SlagleChen, higgs1, higgs2, twisted1, twisted2, fractal, cage} - for a review, see \cite{fractonarcmp}. Fracton phases are quantum phases of matter in which the elementary excitations exhibit {\it restricted mobility}, being either unable to move under local Hamiltonian dynamics, or able to move only in certain directions. This restricted mobility can be traced to the superselection structure of the underlying theories, which in addition to familiar constraints like charge conservation, imposes also additional ``higher moment'' constraints such as conservation of the dipole moment of charge.  Fractons have deep connections with numerous areas of physics, such as elasticity \cite{leomichael,gromov,pai}, gravity \cite{mach}, holography \cite{holo}, quantum Hall systems \cite{ppn}, and deconfined quantum criticality \cite{deconfined}.  Fracton phases are also known to exhibit glassy dynamics.  These have been studied \cite{chamon, KimHaah, SivaYoshida, prem, screening} for fracton systems in three spatial dimensions, using techniques akin to locator expansions. At non-zero densities and long enough times, ``locator expansion'' type techniques fail to converge, likely indicating thermalization at long times. Whether the study of fractons can lead to the discovery of new dynamical quantum phases remains an open question. 

In this work, we combine the techniques of random unitary circuits and ideas from the study of fractons, and discover an entirely new dynamical phenomenon. Specifically, we consider a circuit that conserves not only a $U(1)$ charge (as in \cite{KhemaniVishHuse, Keyserlingk2}), but also the dipole moment of that charge, as an analogue of the ``higher moment'' constraints that underlie the fracton phenomenon. Insofar as energy conservation is essential to immobility of fractons under Hamiltonian dynamics \cite{prem}, and random circuits do not conserve energy, one might have thought that random circuits with fractonic constraints would also thermalize, just like all other random circuit models that have been studied thus far. Instead, we find that in such circuits, fractonic charge {\it fails to spread} even at the longest times - even though the circuit {\it does} contain gates that allow such charge to hop. This unexpected immobility of fractonic charges is traced to the inevitability of returns in low dimensional random walks, akin to the phenomenon of weak localization \cite{AALR}, but immune to dephasing and not limited to the non-interacting limit. A system of coupled hydrodynamic equations is proposed to govern the charge dynamics, and makes several non-trivial predictions which are found to be in good agreement with numerics. Importantly, this system of equations predicts that the localization we observe should persist also to two dimensions, where exact numerical simulations of quantum dynamics are unavailable. At the same time, the stationary fractonic charge emits ``non-conserved operators'', which do spread, but with exponents distinct to those that have been previously observed in random circuits \cite{KhemaniVishHuse, Keyserlingk2}, constituting a new dynamical universality class. These non-standard exponents are also correctly predicted by analysis of our hydrodynamic equations.  The non-ergodic phenomenology is numerically verified to persist in settings with a non-zero density of fractons, including in the sector with close to maximal charge. 

Thus we establish that combining ``fractonic'' constraints with random circuit dynamics can lead to qualitatively new classes of dynamical behavior. At the same time, insofar as random circuit dynamics is a model for noisy time evolution, our results suggest that low dimensional fracton systems should behave as memories forever robust against noise. \footnote{For examples of Hamiltonians realizing fracton phases in two spatial dimensions, see \cite{leomichael, deconfined}. Whether fractons can be realized as a stable phase in one dimensional Hamiltonian systems is still an open question.} Equally remarkably, however, it offers a pathway to breakthrough results in many-body localization. 
 
The remarkable developments in MBL over the past decade have largely relied for analytical understanding on techniques related to locator expansions. A series of works  have established that there exist rare region ``obstructions'' to many-body locator expansions \cite{2dcontinuum, MirlinMuller, QHMBL, Burin, YaoDipoles, deroeck1, deroeck2} in any setting other than strongly disordered one dimensional spin chains. While some of these obstructions can be circumvented \cite{LRMBL, Schwingernumerics, extob} it does appear that systems in dimensions greater than one cannot admit of a fully convergent locator expansion \cite{deroeck1}, and nor can systems with a translation invariant Hamiltonian \cite{deroeck2}. As a result, it is widely believed that higher dimensional systems and/or translation invariant systems can support at best asymptotic or ``quasi'' MBL, in which a memory of the initial conditions survives in local observables for times superpolynomially long in some control parameter \cite{deroeck1}, but the system nevertheless thermalizes on the longest timescales. Our results provide an explicit counterexample to this belief, since they indicate that fractonic systems in one and two dimensions preserve {\it forever} under Hamiltonian time evolution a memory of their initial conditions in local observables, even with translation invariant Hamiltonians. That is, translation invariant one and two dimensional fractonic systems undergoing Hamiltonian dynamics should exhibit {\it true} MBL (not just asymptotic), at least in the sector of Hilbert space with non-zero fracton charge, and this should survive to non-zero energy densities and should be robust to generic local perturbations obeying the fractonic constraint. \footnote{We are defining true MBL in physical terms, as a phase robust against local perturbations that preserves a memory of its initial conditions in local observables for infinite times \cite{mblarcmp}.} Our result evades the ``no-go'' arguments of \cite{deroeck1,deroeck2} because {\it it does not rely on locator expansions} for localization. As such, it provides a first example of MBL beyond locator expansions, and may open a new chapter in the exploration of that field.

The paper is structured as follows. In Sec. II we introduce the basic model that will be studied. In Sec. III we present the results of numerical simulations indicating the lack of spreading of fractonic charge. In Sec.IV we present the basic analytical understanding of random circuit dynamics with fractonic constraints, and test it against numerical simulations. In Sec. V we discuss the implications of our results, and conclude. 

\section{The spin-1 random unitary circuit model}
We consider a one-dimensional chain of $L$ sites, each of which has a spin-1 degree of freedom. The gates in the random unitary circuit at each step in the time-evolution are designed to preserve the total $z$ component of the spins ($S_{z}^{\textrm{total}}$), which serves as a conserved $U(1)$ charge, as in Refs. \onlinecite{KhemaniVishHuse, Keyserlingk2}.  Unlike previously studied systems, however, we consider a random unitary circuit which also conserves the total \textit{dipole moment} ($\vec{P}_{\textrm{total}}$) of this effective charge.  As such, an isolated charge ($i.e.$ a state with an $S_z=1$ in a background of $S_z=0$) cannot propagate without creating additional dipolar excitations - and as such functions as a fractonic charge. 

The time-evolution is governed by a random unitary circuit (as shown in Figure \ref{fig:randomunitary}) consisting of staggered layers of three-site unitary gates. The time evolution unitary is given by 
$U(t) = \prod_{t'=1}^{t}U(t',t'-1)$, where 
\begin{equation}
  U(t',t'-1)=\begin{cases}
    \prod_{i}U_{3i,3i+1,3i+2} & \text{if $t'$ mod $3 = 0$}\\
    \prod_{i}U_{3i-1,3i,3i+1} & \text{if $t'$ mod $3 = 1$}\\
    \prod_{i}U_{3i-2,3i-1,3i} & \text{if $t'$ mod $3 = 2$}
  \end{cases}
\end{equation}
As a result of the conservation laws, each three-site unitary gate $U_{i,i+1,i+2}$ is a $27 \times 27$ block-diagonal matrix, with block structure as shown in Figure \ref{fig:randomunitary}. Each block is a Haar-random unitary of the appropriate size, and is chosen independently of other blocks. There are four $2 \times 2$ blocks in the unitary matrix, corresponding to the following charge- and dipole-conserving qudit flips: $+ - + \leftrightarrow 0 + 0$, $+ - 0 \leftrightarrow 0 + -$, $- + 0 \leftrightarrow 0 - +$, $- + - \leftrightarrow 0 - 0$. The remaining charge configurations have no nontrivial qudit flips.  Despite the small number of gates, this form of time-evolution is the most generic unitary evolution consistent with conservation of charge and dipole moment.  As such, any other quantity which is not conserved by design can change under the action of the circuit, indicating that charge and dipole moment are the only two conserved quantities in the system.  Thus the system is not a conventionally integrable model, which would have an extensive number of conserved quantities.
\begin{figure}[b]
 \centering
 \includegraphics[scale=0.35]{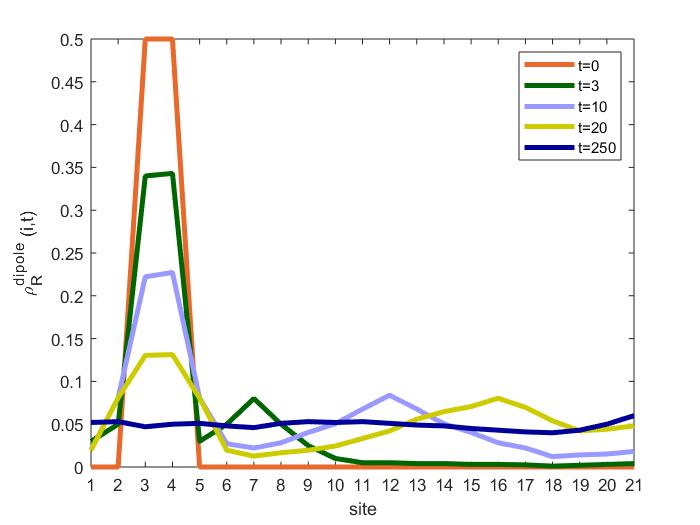}
 \caption{Right-weight profile (averaged over $10$ runs) of the spreading dipole operator in a system of $L=21$ sites with open boundary conditions, at a range of times.}
 \label{fig:dipoleRprofile}
 \end{figure}

In the Heisenberg picture, a local operator $O_{o}$ evolves under $O_{o}(t) = U^{\dagger}(t)O_{o}U(t)$, and the  spatial region over which it has support grows with time. A convenient onsite basis of operators used in the description of the time-evolution of local operators is given by $9$ matrices $\bigg(\Sigma_{i}^{0,1,...,8} = \frac{\lambda_{i}^{0,1,...,8}}{\sqrt{2}}\bigg)$, i.e. the $3 \times 3$ identity matrix ($\lambda_{i}^{0}$) and the $8$ Gell-Mann matrices ($\lambda_{i}^{1,...,8}$). The onsite basis has the following normalization: $\textrm{Tr}\big[ \frac{\Sigma_{i}^{\mu}\Sigma_{i}^{\nu}}{3}\big]=\delta_{\mu \nu}$. We use these matrices to form a basis of $9^{L}$ generalized Pauli strings $S=\prod_{i}\Sigma_{i}^{\mu_{i}}$. The initially-local operator $O_{o}$ consists only of those strings that are the identity at all sites except a few near the site of initialization. With time, however, there are other basis strings with nonlocal weight in the decomposition of the time-evolved operator $O(t)$. In the string basis, the operator $O(t)$ can be expanded as 
\begin{equation}
O(t) = \sum_{S} a_{S}(t)S
\end{equation}
The basis strings satisfy $\textrm{Tr}[SS']/3^{L} = \delta_{SS'}$, making it possible to obtain the coefficients as follows: $a_{S}(t) = \textrm{Tr}[SO(t)]/3^{L}$. The initial operator $O_{o}$ is normalized such that $\textrm{Tr}[O_{o}O_{o}] =3^{L}$, which implies
that the total weight of $O(t)$ on all strings $S$ is $1$, which is just a statement about unitarity in the dynamics:
\begin{equation}
\sum_{S} |a_{S}(t)|^{2} = 1
\label{unitarity}
\end{equation}

In order to determine expectation values of operator, consider an initial density matrix 
\begin{equation}
\rho(0) = (\mathbb{I}_{\textrm{background}} + cO_{o})/3^{L},
\end{equation}
where $\mathbb{I}_{\textrm{background}}$ is the background state which is initially the identity on the full system; $cO_{o}$ is a traceless local onsite
perturbation at the origin to this initial state. At subsequent time, we take all expectation values with respect to a density matrix $\rho(t) = (\mathbb{I}+O(t))/3^L$.

\section{Characterization of operator spreading}
In order to study the operator spreading profile of an initially-local operator $O_{o}$, we follow Ref. \cite{KhemaniVishHuse} and define the ``right-weight'' $\rho_{R}(i, t)$ as the total weight in $O(t)$ of basis strings that end at site $i$, meaning that they act as the identity on all sites to the right of site $i$, but act as combinations of the identity and non-identity operators upto $i$. This gives
\begin{equation}
\rho_{R}(i, t)= \sum_{\substack{$\small strings$\ S\ $\small with$\\ $\small rightmost$\ $\small non-identity$\\ $\small at$\ $\small site$\ i}} |a_{S}(t)|^{2} 
\end{equation}
From Eq. \ref{unitarity}, we see that there is a conservation law on $\rho_{R}(i, t)$, i.e. $\sum_{i}\rho_{R}(i, t)=1$. This gives $\rho_{R}(i, t)$ the interpretation of an emergent local conserved density, the spreading of which can then be examined.

\subsection{Diffusive spreading of local conserved dipole moment}
As a warmup, we first consider the case where the initial operator is a two-site conserved dipole operator, i.e. $O_{o} = \frac{1}{2}\big( \otimes_{i=1}^{o} \mathbb{I} \otimes S_{z} \otimes_{i=o+2}^{L} \mathbb{I} - \otimes_{i=1}^{o-1} \mathbb{I} \otimes S_{z} \otimes_{i=o+1}^{L} \mathbb{I} \big)$, 
where $S_{z}$ is given by the matrix
\begin{equation}
S_{z} = \begin{pmatrix}
1 & 0 & 0 \\
0 & 0 & 0\\
0 & 0 & -1
\end{pmatrix}
\label{Szmatrix}
\end{equation}
The spin-1 chain (initialized with the above local dipole operator $O_{o}$) is time-evolved under the action of the random unitary circuit. We evaluate the exact Heisenberg time evolution of the operator by multiplying it by the appropriate unitary gates, a procedure that is even simpler than exact diagonalization. A further speedup results because the constraints on the unitary operators allow us to restrict to a single charge and dipole sector, such that one does not need to keep track of all $9^{L}$ (where $L$ is the system size) coefficients at each time-step. This is repeated for various values of $t$ to obtain the right-weight profile as in Figure $2$.

\begin{figure}[b]
 \centering
 \includegraphics[scale=0.35]{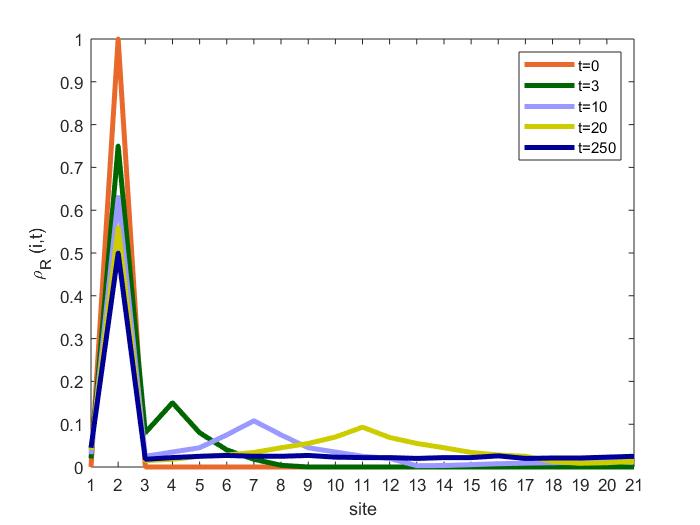}
 \caption{Right-weight profile (averaged over $10$ runs) of the spreading fracton charge operator in a system of $L=21$ sites with open boundary conditions.}
 \label{fig:chargeRprofile}
 \end{figure}
 
Since the initial condition involves a dipole operator, which acts as a conserved $U(1)$ charge without any conservation laws on its higher moments, this is a situation that should map onto the analysis of \cite{KhemaniVishHuse}, and serves as a sanity check on our results. As discussed in \cite{KhemaniVishHuse} the spatial structure of this operator shows a ballistic front, a power-law tail behind the front, and diffusively spreading charges near the origin. 
%
This is illustrated in Figure \ref{fig:dipoleRprofile}, which shows the right-weight profile $\rho^{\text{dipole}}_{R}(x, t)$ for a system of size $L=21$ initialized with a single conserved dipole. These profiles depict the following regimes: a slowly spreading ``lump'' at the origin, a ``front'' that moves out rapidly, and a ``tail'' behind the front. In subsequent sections we show that the front propagation is ballistic, and that the exponent governing the tail matches the analysis in \cite{KhemaniVishHuse}. It is also apparent that in the long time limit, the central lump relaxes to a uniform charge distribution. These results, consistent with \cite{KhemaniVishHuse}, serve as a sanity check on our procedure. We do not discover anything new simply by considering an initial condition with a single conserved dipole. To obtain new results, we must consider initial conditions containing {\it fractons}, a task to which we turn in the following subsection. 

\begin{figure*}[t]
\centering
\begin{subfigure}{0.99\columnwidth}
\includegraphics[width=\columnwidth]{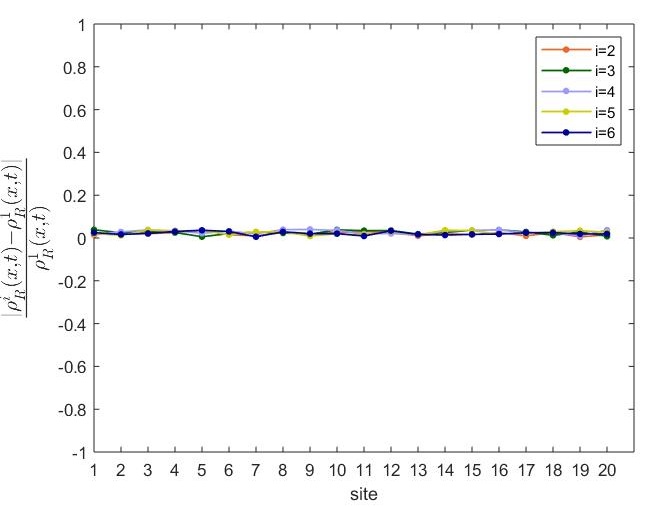}%
\caption{Circuit to circuit fluctuations in the right-weights of the fracton charge operator for six different runs computed at late times. $i$ in $\rho^{i}_{R}$ labels typical runs.}%
\label{circuitfluctuations}%
\end{subfigure}\hfill%
\begin{subfigure}{0.9\columnwidth}
\includegraphics[width=\columnwidth]{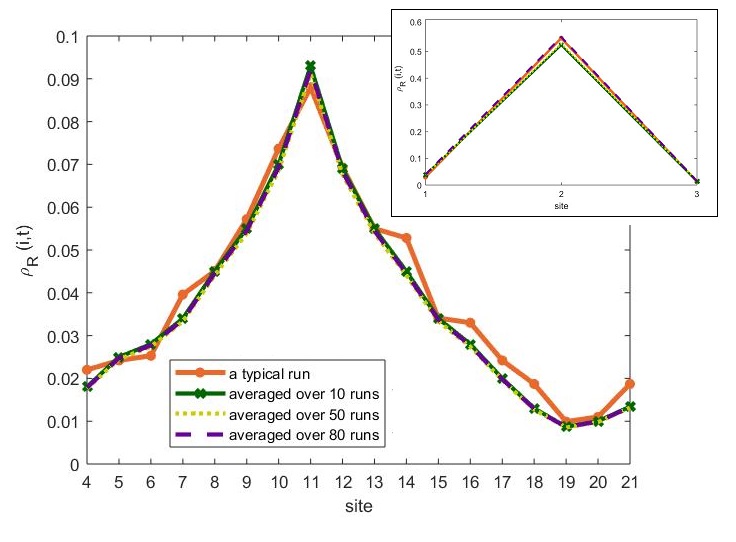}%
\caption{Right-weight profile of the fracton operator averaged over different number of disorder realizations. The right-weight profile for sites $4$ through $21$ is shown in the plot and that for sites 1-3 is shown in an inset.  This demonstrates that averaging over $\mathcal{O}(10)$ configurations is sufficient. }%
\label{disorderaverage}%
\end{subfigure}\hfill%
\caption{Fluctuations in the right-weights of the fracton charge operator.}
\label{fig:fluctuations}
\end{figure*}

\noindent

\subsection{Spreading of conserved fracton charge : memory of initial conditions}
We now consider the case where the initial operator is a conserved fractonic charge, i.e. $O_{o} = \otimes_{i=1}^{o-1} \mathbb{I} \otimes S_{z} \otimes_{i=o+1}^{L} \mathbb{I}$. Now the situation is unlike \cite{KhemaniVishHuse}, since not only is the monopole moment of the initial operator conserved, but so is its \emph{dipole moment}. The resulting behavior is strikingly altered, as is illustrated in Figure \ref{fig:chargeRprofile}. Unlike the dipole right-weight profile, the spatial structure of this operator shows an extremely prominent peak where the operator was initially localized, i.e. \emph{the system has memory of its initial conditions in a local observable} at all times accessible in our simulations. Additionally, there is a ``propagating front'' with a lagging tail and, as we shall show in a subsequent section, the power law governing this lagging tail is distinct to the power law observed in Ref.\cite{KhemaniVishHuse} and Figure \ref{fig:dipoleRprofile}. It is thus clear that the evolution of an initial fracton operator lies in a different universality class. 



An important point to note is that circuit to circuit fluctuations in the right-weight profiles of both dipole and fracton charge operators are negligible. We show this in Figure \ref{fig:fluctuations} for the case of the fracton charge operator by plotting the absolute value of the difference in the right-weights for six typical runs relative to one of the runs. We also show in the Appendix that these results hold true for a spin-$S$ ($S \neq 1$) chain. 

\begin{figure*}[t]
\centering
\begin{subfigure}{0.99\columnwidth}
\includegraphics[width=\columnwidth]{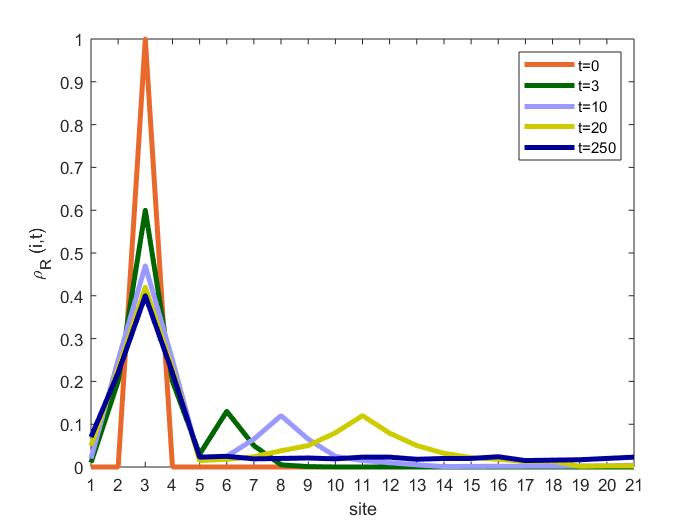}%
\caption{$L=21$ spin chain with 4-qudit gates}%
\label{4qudit}%
\end{subfigure}\hfill%
\begin{subfigure}{1.0\columnwidth}
\includegraphics[width=\columnwidth]{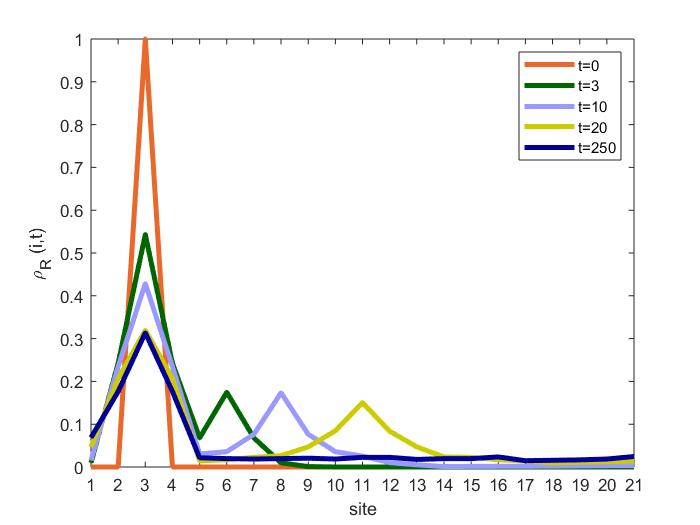}%
\caption{$L=21$ spin chain with 5-qudit gates}%
\label{5qudit}%
\end{subfigure}\hfill%
\caption{Right-weight profile (averaged over $10$ runs) of the spreading fracton charge operator in a system of $L=21$ sites with gates of different sizes (with open boundary conditions).}
\label{fig:gatesize}
\end{figure*}

The persistent memory of the initial condition is remarkable, since the circuit does contain gates that allow a fracton to move (by creating dipole excitations) - see Table \ref{qudit flip gates}. One might worry that this might be an artefact of our choice to limit the circuit to three qudit gates, which block diagonalize into at most two by two blocks. To test for this possibility, we look at what happens to the fracton operator peak if we use gates of different sizes, which allow many more transitions. We consider the cases involving 4- and 5-qudit gates (see Table \ref{qudit flip gates}), and in Figure \ref{fig:gatesize} we see that the fracton peak still remains.  While larger gate size results in a broader peak at large $t$, the integrated area under the fracton peak with respect to the background value at large $t$, i.e. $\omega_{\text{peak}}(t)-\omega_{\text{peak}}(\infty)$,  is independent of gate size and approaches its asymptotic value as $t^{-3/2}$, as shown in Figure \ref{fig:gatesizeintarea}. We explain this exponent in Section \ref{operator growth}.

\begin{figure}[b]
 \centering
 \includegraphics[scale=0.55]{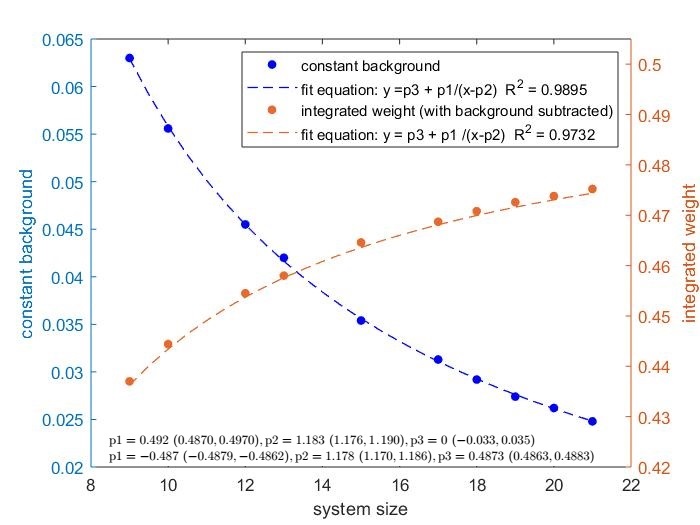}
 \caption{Integrated weight of fracton peak in the right-weight as a function of system size at large $t(=200)$ (averaged over $10$ runs). The uncertainties in the fitting parameters are given below the plot.}
 \label{fig:intweight}
 \end{figure}

\begin{figure}[b]
 \centering
 \includegraphics[scale=0.35]{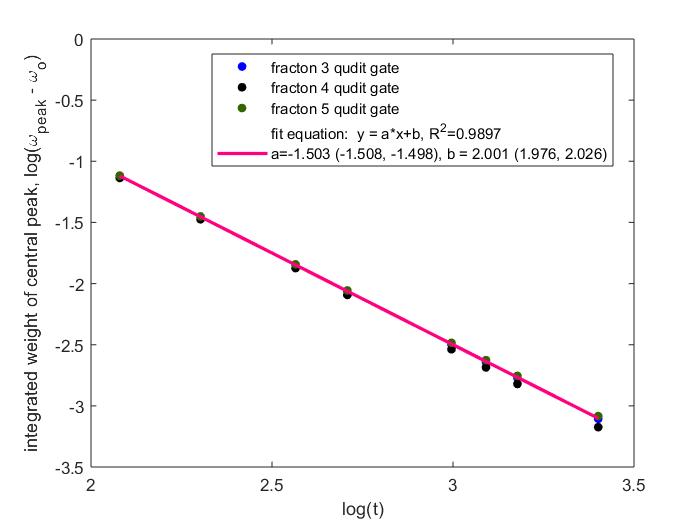}
 \caption{Log-log plot of the integrated area under the fracton peak in the right-weight plots ($\omega_{peak}$) as a function of time for three different gate sizes (averaged over $10$ runs). On the $y$ axis, we plot the logarithm of the area under the central peak at time $t$ minus the area under the central peak at $t \rightarrow \infty$. }
 \label{fig:gatesizeintarea}
 \end{figure}

As an additional check on the robustness of this localization, we also plot (Figure \ref{fig:intweight}) the constant background of the right-weight profile (seen at large $t$) and the integrated weight under the fracton peak as a function of system size. As can be seen from the plot, the weight under the fracton peak persists in the thermodynamic limit, making it safe to attribute this behavior to the fractonic constraints in the system.

\begin{figure*}[t]
  \begin{subfigure}{8cm}
    \centering\includegraphics[width=8.3cm]{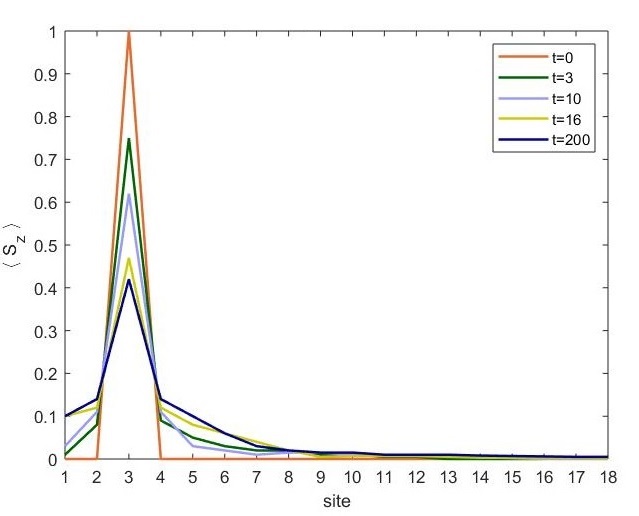}
    \caption{$\langle S_{z} \rangle$ vs position for different values of $t$ in an $L=18$ spin chain.}
  \end{subfigure}
  \begin{subfigure}{8cm}
    \centering\includegraphics[width=8.3cm]{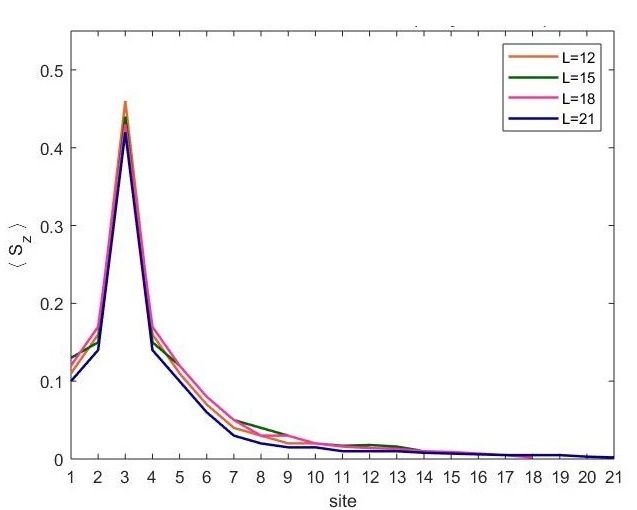}
    \caption{$\langle S_{z} \rangle$ vs position for different system sizes at large $t\ (=200)$}
  \end{subfigure}
  \caption{$S_{z}$ expectation values (averaged over $10$ runs) in the spin-1 random unitary circuit model, initialized with a single fracton.}
   \label{fig:SzExpValue}
\end{figure*}

\begin{figure}[b]
 \centering
 \includegraphics[scale=0.50]{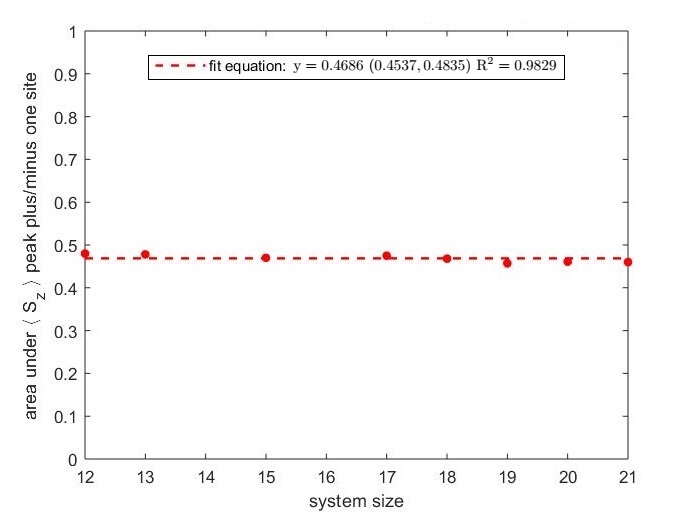}
 \caption{Area under the $\langle S_{z} \rangle$ peak as a function of system size at large $t (=200)$ (data averaged over $10$ runs). The peak is defined as the region bounded by central site $\pm 2$.  The uncertainty in the fitting parameter is given in the legend.}
 \label{fig:intareaSz}
 \end{figure}

Thus far we have been considering the ``emergent'' hydrodynamics of unitary operator spreading (translated into constraints on right-weight profiles). To confirm that the memory of initial conditions is manifest in the dynamics of the physical components of the system, we look at what the charges themselves are doing, and not just the right-weights. We examine what happens to $\langle S_{z} \rangle$ as a function of position. The ``fracton peak'' still remains, and this is largely independent of time and system size, as can be seen from the plots in Figure \ref{fig:SzExpValue}. A plot of the integrated area under the $\langle S_{z} \rangle$ peak as a function of system size at large $t$ (Figure \ref{fig:intareaSz}) shows that there is considerable operator weight that remains at the original location, i.e. the system acts as a memory which is robust against random noise. Note the absence of any spreading front in this plot: fractonic charge does not spread under the action of the random circuit, even though there are gates that apparently allow charge to move.

\begin{figure*}[t]
  \begin{subfigure}{8cm}
    \centering\includegraphics[scale=0.47]{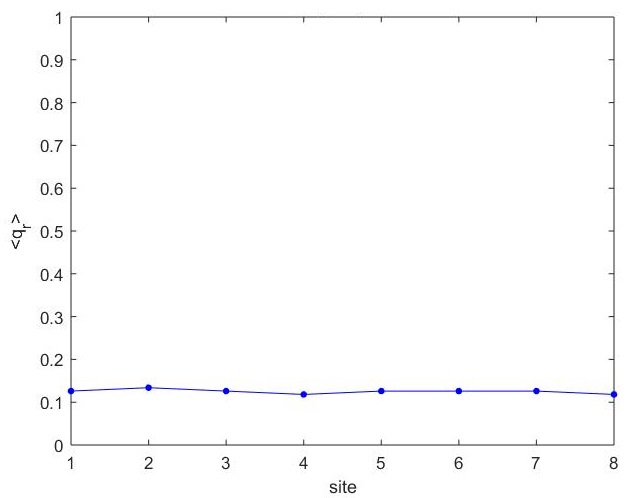}
    \caption{$Q=1, P=2, L=8$}
  \end{subfigure}
  \begin{subfigure}{8cm}
    \centering\includegraphics[scale=0.47]{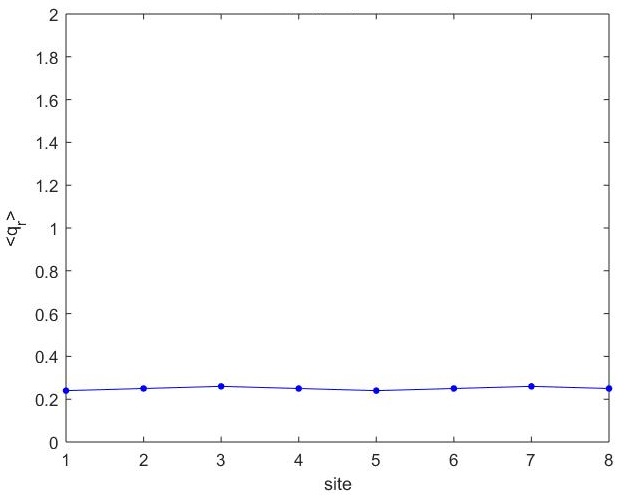}
    \caption{$Q=2, P=2, L=8$}
  \end{subfigure}
  \caption{$\langle S_{z} \rangle$ in the numerically obtained thermal/maximal entropy state of a classical $3$-state model with fixed charge $Q$ and dipole moment $P$. The above plot is for periodic boundary conditions, with dipole moment defined mod $L$ (the choice of origin is arbitrary, and the anticlockwise direction is taken to be positive). The results for open boundary conditions are similar: there is no localization of charge in the ``classical'' maximum entropy state.}
   \label{fig:thermalstate}
\end{figure*}

Importantly, the peaked late-time $\langle S_z\rangle$ profile seen in Figure \ref{fig:SzExpValue} is demonstrably different from the thermal state with the same charge and dipole moment.  The thermal state should simply be the state of maximal entropy.  To determine what $\langle S_{z} \rangle$ looks like for the maximal entropy state, we consider a \emph{classical} three state model on a length $L$ chain, where every site has charge $q = +1, 0,$ or $-1$. We then consider the set of all classical states that satisfy the constraints on charge and dipole moment, \emph{i.e.} $\sum_i Q_i = \mathcal{Q}$, and $\sum_j j Q_j = \mathcal{P}$. We then construct $\rho$, the uniform weight superposition of all density matrices in this set. If we look at $\textrm{tr}(\rho Q_{r})$, we see that it is spatially uniform (see Figure \ref{fig:thermalstate}) in any sector $(\mathcal{Q}, \mathcal{P}, L)$ ($\mathcal{Q}=$total charge, $\mathcal{P}=$ total dipole moment, $L=$system size) \footnote{This figure is for periodic boundary conditions, such that dipole moment is only defined mod L. However, similar results are obtained with open boundary conditions}. While the maximum entropy state has a \emph{flat} $\langle S_{z} \rangle$ distribution, the state that we obtain from our random circuit model has a prominent peak in the $\langle S_{z} \rangle$ profile (Figure \ref{fig:SzExpValue}). Therefore, the system at late times is indeed \emph{non-thermal}.

\section{Operator growth in the fractonic circuit}

We now develop a basic analytic understanding of the various features uncovered in the previous sections, which we test against numerics.  In the case with only local charge conservation (and no dipole conservation) \cite{KhemaniVishHuse}, the charge executes a random walk under the action of a random unitary circuit. In our model likewise, the dipoles execute random walks under the action of a random unitary circuit. When we consider a system having fractons, however, the dipole conservation constraint means that these can only move by absorbing/emitting a dipole (eg., $+ - + \leftrightarrow 0 + 0$). At first glance, one might think that since gates exist that move fractons (at the cost of creating/annihilating dipoles), the fractons should also execute a random walk under the action of the random circuit. However, there is a crucial distinction: the fracton remains displaced only if the dipole stays absorbed/emitted. If a fracton \emph{moves} by emitting a dipole, then a subsequent reabsorption of that dipole will return the fracton to its initial position. 

The dipoles undergo a random walk. It is well known that random walks in spatial dimension $d \le 2$ always return to the origin.  This can be understood in terms of the diffusion propagator, $G(x,t) = (4\pi Dt)^{-d/2} e^{-x^{2}/Dt}$, where $D$ is diffusion constant of the dipoles, and $d$ is the spatial dimension. In order to determine the return probability  of the dipole to its starting point, we consider the following integral:
\begin{equation}
\int dt\ G(0,t) = \int dt\ \bigg(\frac{1}{\sqrt{4\pi Dt}}\bigg)^{d}  
\end{equation} 
In $d=1$, this integral diverges ($\sim \sqrt{t}$), therefore the probability that the dipole returns to its initial position is $1$, i.e. the dipole always diffuses back to its starting point.  When the dipole returns, this causes the fracton to return to its initial position. Thus, although the fracton can temporarily hop at any given time step, the dipole \emph{always eventually diffuses back}, returning the fracton to its original position. This explains the persistent peak that we see in the right-weight and $\langle S_{z} \rangle$ profiles even as $t \rightarrow \infty$. 
 
This simple argument suggests that even though fractonic circuits contain gates that move fractons, nevertheless fractons should be immobile under random circuit dynamics not only in $d=1$ (which we have studied numerically), but also in $d=2$ (which is beyond reach of our numerical techniques). In $d=3$, however, the behavior will qualitatively change, since three-dimensional random walks do not inevitably return to the origin, and dipoles can actually escape to infinity. So in $d=3$, fractons should be able to execute a true random walk via permanent emission of dipoles. This suggests that the behavior of fractonic circuits in $d=3$ is very different from that in $d=1,2$. Namely, in $d=3$ a single fracton evolving under a random unitary circuit should be able to diffuse, but not in $d=1,2$.  

These arguments suggest that the fractonic circuit should be well-modeled by a picture of a conserved dipole density coupled to a mobile source/sink (the fracton). The hydrodynamic equations that model the system should then take the form
\begin{align}
\begin{split}
\frac{d\vec{R}}{dt} &= - \vec{\eta} + \vec{A}(t) \\
\frac{d \vec{\eta}}{dt} &= D \nabla^2 \vec{\eta} + \frac{d\vec{R}}{dt}\ \delta(\vec{x}-\vec{R}),
\end{split}
\label{Hooke}
\end{align}
where $\vec{\eta}$ is the \emph{local} dipole charge density at the location of the fracton, $D$ is the dipole diffusion constant, $\vec{A}(t)$ is a stochastic force (taken to be delta-correlated white noise), and $\vec{R}$ is the position of the fracton ($i.e.$ the location of the single isolated $S_z=1$ state). In one spatial dimension, the quantities $\vec{R}, \vec{\eta}, \vec{A}$ are all scalars. In higher spatial dimensions, they will be vector quantities. We have written the equations in terms of vector quantities for maximum generality. These equations are expected to describe equally well the system in any spatial dimension, but (as we will show), the character of the solutions is highly sensitive to spatial dimension. We now explore the predictions of this simple picture, and test it against numerics in one spatial dimension. 

\begin{figure}[b]
 \centering
 \includegraphics[scale=0.55]{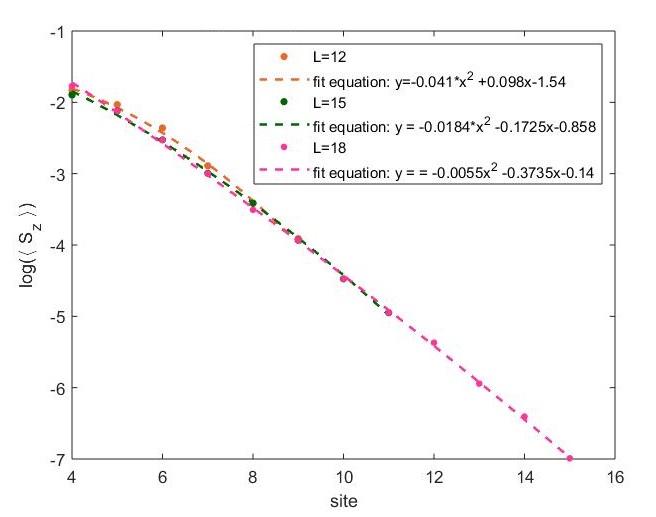}
 \caption{The fracton charge probability distribution function is given by the Boltzmann distribution: the tails of $\langle S_{z} \rangle$ have been plotted on a semi-log scale for different system sizes ($L$) as a function of position at large $t$ (averaged over $10$ runs).}
 \label{fig:semilogSz}
 \end{figure}

\subsection{Fracton localization and sensitivity to dimensionality}
We now solve the system of equations introduced above, and show that they imply fracton localization in one and two spatial dimensions (but not in three dimensions). We initialize the problem with a single fracton at position zero, and zero dipole charge. The total dipole charge $\tilde {\vec{\eta}}$ is related to the position of the fracton $\vec{R}(t)$ via 
\begin{equation}
\tilde{\vec{\eta}}(t) =  \vec{R}(t),
\label{dipoledensity}
\end{equation}
However, what enters into Eq.(\ref{Hooke}) is not the {\it total} dipole charge, but rather the dipole charge density at the position of the fracton. To estimate this, we need to estimate how much the dipole density \emph{spreads out}. We can do this as follows. First, we estimate a typical ``return time'' (i.e. the typical time that a dipole wanders freely before it returns to the fracton to get reabsorbed). This can be estimated from the equation $\int_1^t G(0,t') dt' = C$ where C is some arbitrary large but finite constant, and $G(x,t)$ is the diffusion propagator. This gives 
\begin{eqnarray}
t_{return} &=&  \pi D C^2 \qquad \qquad d=1 \nonumber\\
&=& \exp(4 \pi D C) \qquad d=2 \nonumber\\
&=& \infty \qquad \qquad \qquad d>2
\end{eqnarray}
where $D$ is the dipole diffusion constant. The length scale $\xi$ over which the dipole charge density is spread out can then be estimated as $\xi \sim \sqrt{D t_{return}}$ (being cutoff in $d>2$ by the scale of the system size $L$).  The density at the fracton position can then be estimated as $\tilde{\vec{\eta}}/\xi^d$. Substitution into Eq.(\ref{Hooke}) then gives an equation of motion for the fracton of the form
\begin{equation}
\frac{d\vec{R}}{dt} = - \frac{1}{\gamma} \vec{R} + A(t) \label{Langevin}
\end{equation}
where 
\begin{eqnarray}
\gamma &\approx &\sqrt{\pi} D C \quad \qquad \qquad d=1 \nonumber\\
&\approx& D\exp(4 \pi D C) \qquad d=2 \nonumber\\
&\approx& L^d \qquad \qquad \quad \qquad d >2
\end{eqnarray}
Now Eq.(\ref{Langevin}) can just be recognized as a Langevin equation, which has the well-known solution 
\begin{equation}
P(\vec{R}) \sim \exp \left(-\frac{R^2}{2 \gamma T} \right), \label{locsteadystate}
\end{equation}
where $P(R)$ is the probability distribution for the fracton position, and $T$ is the strength of the stochastic kick, $\langle A(t) A(t')\rangle = 2 T \delta(t-t')$. Now, in $d=1,2$ the coefficient $\gamma$ is finite and independent of system size, so this predicts exponential localization of the fractons in spatial dimensions $d=1,2$. Meanwhile, in $d=3$, $\gamma \sim L^d$ so the ``localization length'' is greater than system size, and thus the fracton is effectively spread over the whole system in the steady state. 

\begin{figure}[b]
 \centering
 \includegraphics[scale=0.35]{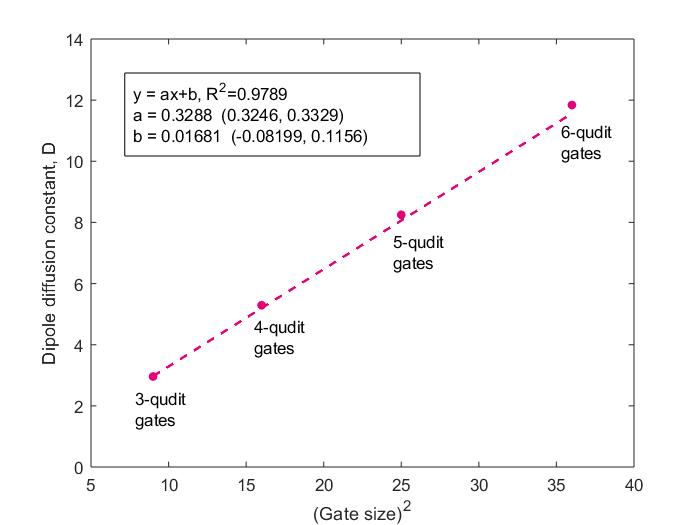}
 \caption{Plot showing how the diffusion constant of dipoles scales with gate size. The diffusion constant is extracted by preparing an initial condition with a single dipole, tracking how dipole charge density at the origin decays as a function of time, and fitting to $1/\sqrt{Dt}$. We obtain a behavior $D \sim (\textrm{gate size})^2$, consistent with our expectations. \label{diffusionconstantgatesize} }
\end{figure}

Thus, solving the system of hydrodynamic equations yields the prediction that fracton charge should be localized in one and two dimensions, but not in three dimensions. Note also that the localization length in two dimensions is considerably larger than in one dimension (exponentially large in the diffusion constant of the dipoles, instead of power law large). 

\begin{figure}[t]
\centering
\includegraphics[scale=0.35]{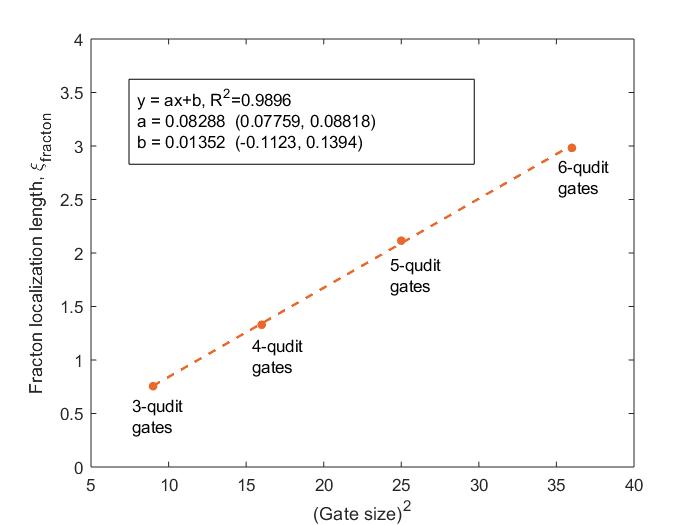}
\caption{Plot showing how the charge localization length $\xi_{\textrm{fracton}}$ scales with gate size. The localization length is obtained by looking at $\langle S^z(r)\rangle$ in the long time steady state, and is defined simply as the half width at half maximum of the $S^z$ peak. We find a scaling $\xi_{\textrm{fracton}} \sim (\textrm{gate size})^2$, consistent with the predictions of our analysis. \label{loclengthgatesize}}
\end{figure}

We now compare the predictions of this analysis to direct numerical simulation in one dimension. Figure \ref{fig:semilogSz} shows a semi-log plot of the tails of $\langle S_{z} \rangle$ as a function of site position at large $t$, and the exponential behavior seen is consistent with the above analysis, although we are not able to distinguish between $\exp(-x)$ and $\exp(-x^2)$ behavior numerically. 
\begin{figure*}[t]
  \begin{subfigure}{8cm}
    \centering\includegraphics[scale=0.32]{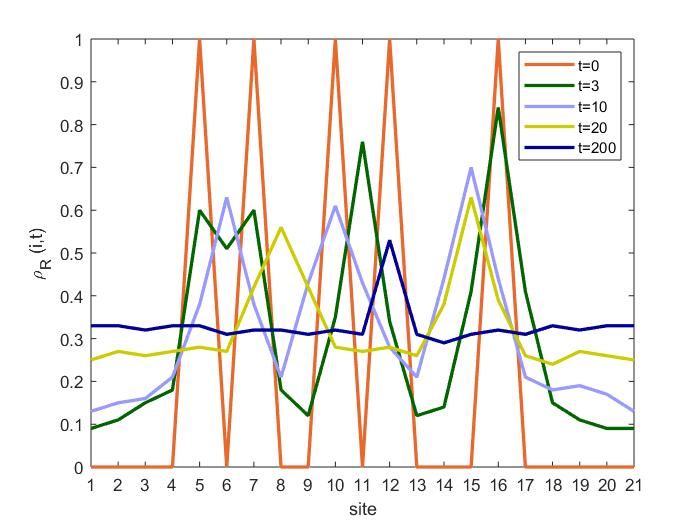}
    \end{subfigure}
  \begin{subfigure}{8cm}
    \centering\includegraphics[scale=0.32]{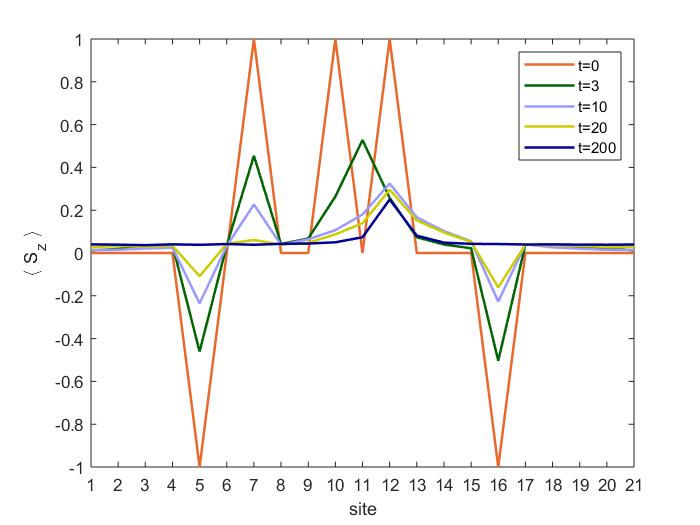}
  \end{subfigure}
  \caption{A fracton initialized at site $10$ in an $L=21$ system (under periodic boundary conditions) with a strength $-4$ dipole on its right and a $+2$ dipole on its left shifts to the right by $2$ sites, i.e. $\tilde{\eta_{0}} = -2$ in this case. (Data from a typical run).}
   \label{fig:brownian}
\end{figure*}

A nontrivial prediction coming from our analysis of the hydrodynamic equations is that the localization length for fractonic charge should be sensitive to the diffusion constant of the dipoles. This sensitivity is particularly strong in $d=2$, where the localization length is exponential in the dipole constant, but even in $d=1$ we predict that the localization length for charge should scale as $\xi_{\textrm{fracton}} \sim \sqrt{D T}$. We now test this prediction directly against numerics. 

We remind the reader that our model of quantum dynamics is minimally structured, \emph{i.e.} we are considering a {\it random} circuit, made out of gates of a particular size satisfying the charge and dipole moment constraint. Once the gate size is fixed, the parameters $D$ and $T$ are fixed also. The only way to change these parameters is to change gate size. Gate size translates directly into ``step size'' for the dipoles and fractons, and we therefore expect that $D \sim (\textrm{gate size})^2$ and also $T \sim (\textrm{gate size})^2$ (remember $T$ is just the correlator of the stochastic steps taken by the fractons). This expectation can be straightforwardly checked in numerics, by varying the gate size and looking at the spreading of dipole charge in an initial condition containing a dipole only (see Figure \ref{diffusionconstantgatesize}) and indeed we find the expected behavior. We therefore predict that charge localization length $\xi_{fracton} \sim \sqrt{DT} \sim (\textrm{gate size})^2$, where the charge localization length could be extracted simply as the half width at half maximum on a plot of $\langle S^z \rangle$ as a function of position. This is a prediction that can be directly tested against numerical simulations, and we find excellent agreement with our prediction (see Figure \ref{loclengthgatesize}).
 
Finally, the analysis can be straightforwardly adapted to initial conditions containing a non-zero dipole charge $\tilde{\vec{\eta_0}}$. We can think of this initial condition as having ``descended'' from an initial condition with no dipoles, where the fracton started at position $-\tilde{\vec{\eta_0}}$. The steady state will then be obtained simply by shifting (\ref{locsteadystate}), and will take the form 
 \begin{equation}
P(\vec{R}) \sim \exp \left(-\frac{|\vec{R}+\tilde{\vec{\eta_0}}|^2}{2 \gamma T} \right) \label{shiftedlocalization}
\end{equation}
 i.e. in steady state the fracton shifts position by $-\tilde{\vec{\eta_0}}$, thereby absorbing the excess dipole density, but remains localized about this shifted position. This prediction can also be tested numerically. For simplicity we work with periodic boundary conditions, such that there is only a single dipole charge characterizing the system. (Here, left and right are defined with respect to an arbitrary choice of origin, with the anticlockwise direction being ``right'', and dipole moment is defined mod $L$). Our solution of the governing hydrodynamic equations (\ref{shiftedlocalization}) predicts that for an initial condition with a fracton and dipole charge $\eta_0$, at late times, the fracton peak will be found shifted left by $\eta_0$ i.e. the fracton will end up absorbing all the dipoles. This basic picture is confirmed by direct numerical simulations Figure \ref{fig:brownian}. Note that, in configurations such as Figure \ref{fig:brownian}, the identification of which charges make up dipoles versus individual fractons is inherently ambiguous.  However there is no ambiguity in the location of the final peak.

We thus conclude that our analysis of the coupled hydrodynamic equations (Eq.\ref{Hooke}) makes a number of nontrivial predictions which are all in agreement with numerics in $d=1$, where high quality numerical tests are possible. As we will show, these equations also accurately describe the universal dynamics of the spreading non-conserved part of the operator (next subsection). This excellent agreement between analytics and numerics gives us high confidence in our analysis, even in two dimensions, where direct numerical simulation is not possible. We recall that our analysis predicts localization in two dimensions (but not in higher dimensions). 

\subsection{Propagating fronts and power law tails}
\begin{figure}[t]
 \centering
 \includegraphics[scale=0.3]{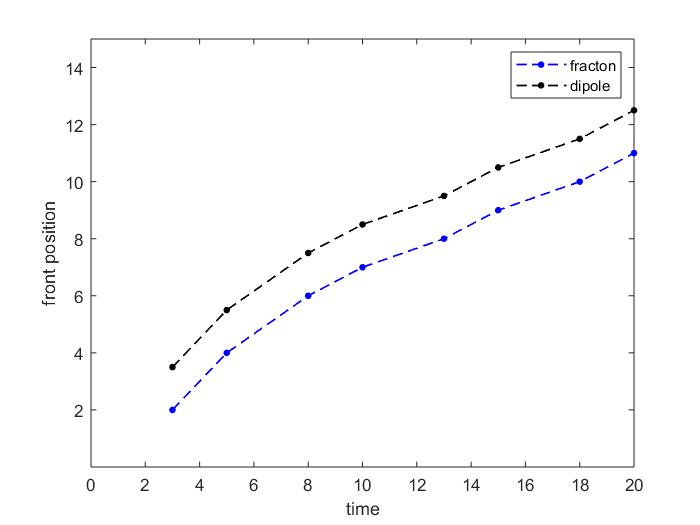}
 \caption{Position of the propagating front as a function of time in the spin-1 model for the fracton  and dipole operators  ($L=21$) for a typical run.}
 \label{fig:frontposition}
 \end{figure}

\begin{figure}[b]
 \centering
 \includegraphics[scale=0.35]{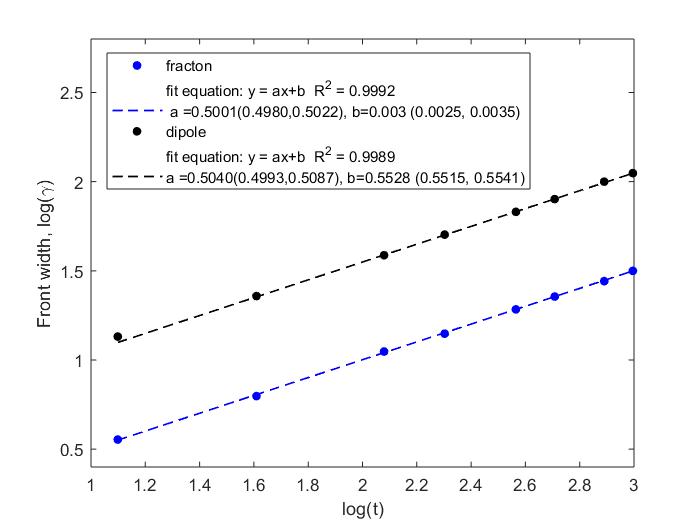}
 \caption{Front broadening in the right-weight profiles (averaged over $10$ runs) of the fracton and dipole operators ($L=21$). The uncertainties in the fitting parameters are given in the legend.}
 \label{fig:front}
 \end{figure}

For either an initially localized fracton or dipole, the operator spreading profile contains a ballistically propagating front.  The front itself behaves in the same way in the fracton and dipole cases, $i.e.$ it propagates ballistically with a similar velocity, featuring front broadening that shows a power-law exponent of $1/2$ when plotted against time $t$. This is due to the fact that the front consists of nonconserved operators emitted at the earliest times and is not influenced by what happens near the origin at late times. This behavior is numerically verified in the plots in Figure \ref{fig:frontposition} and Figure \ref{fig:front}. 

\begin{figure}[t]
 \centering
 \includegraphics[scale=0.57]{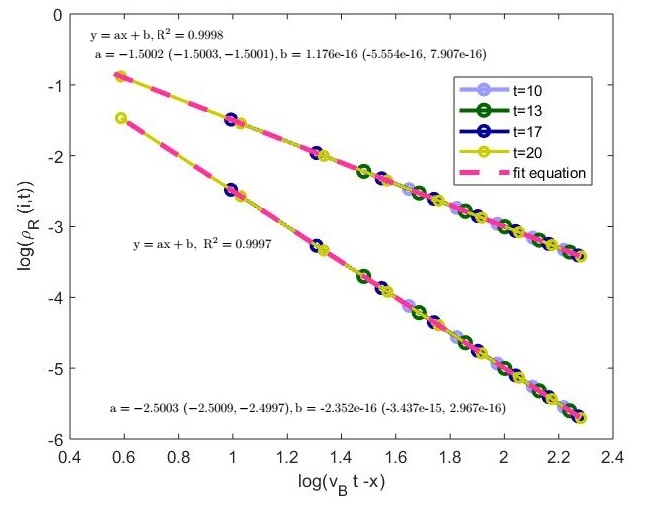}
 \caption{Power-law behavior of diffusive tails behind propagating fronts ($L=21$). The right-weight profiles (averaged over $10$ runs) have been rescaled before fitting. The dipole operator has power-law exponent $-3/2$, while the fracton operator has an anomalous exponent of $-5/2$. Fitting parameter uncertainties are specified in the plot.}
 \label{fig:tails}
 \end{figure}

In addition to the ballistic front in the operator spreading profile, there is important information contained in the structure of the tail behind the front, which behaves {\it differently} in the cases of dipole and fracton charge operators (see Figure \ref{fig:tails}). We observe that the diffusive tails show power-law behavior in both cases and extract the exponents.  In the case of the dipole operator, the tails show power-law behavior of the following form: $(v_Bt-x)^{-3/2}$, where $v_B$ is the velocity of the ballistic front. This is consistent with the discussion in Ref. \cite{KhemaniVishHuse} and can be understood analogously, $i.e.$ the nonconserved parts of the operator are sourced by the diffusing charge, with strength $\sim d\rho_{c}/dt$, where $\rho_c$ is the weight in the conserved charge sector. But this weight goes as $\rho_c \sim 1/\sqrt{t}$ in one dimension, so this naturally gives the tails due to the nonconserved parts emitted at later times the form $d\rho_{c}/dt\sim t^{-3/2}$. 
 \label{operator growth}
For the case of an initially localized fracton operator, the tails satisfy power law behavior too, but with exponent $-5/2$, i.e. the tails have the form  $(v_Bt-x)^{-5/2}$.  This exponent can also be naturally explained in terms of emission of nonconserved operators from the dipole sector, which in turn is stochastically sourced/sinked by the fracton. That is, fracton motion will generate dipole density, which will then be governed by diffusion close to a sink (the fracton itself, which can reabsorb the dipole).  The problem of diffusion close to a sink has been studied previously \cite{notes}, and it is known that, close to the sink, the resulting diffusion propagator is given by a spatial derivative of the free-space diffusion propagator:
\begin{equation}
G(x,t) \sim \frac{d}{dx}(t^{-d/2}e^{-x^2/Dt})\sim \frac{x}{t^{1+(d/2)}}e^{-x^2/Dt}
\end{equation}
which gives the \emph{amplitude} of a dipole to be at a specific time and place.  The total \emph{weight} in the central peak is then given by:
\begin{equation}
\rho_c = \int_0^\infty dx |G(x,t)|^2\sim t^{-(2d+1)/2}
\end{equation}
We see that the weight in the conserved sector decays faster, as $\rho_c \sim t^{-3/2}$ (consistent with our prior observation in Figure \ref{fig:gatesizeintarea}), and thus by an argument analogous to \cite{KhemaniVishHuse} produces tails that go like $d\rho_{c}/dt \sim t^{-5/2}$. \footnote{It is interesting to note that in the absence of dipole conservation \cite{KhemaniVishHuse}, an initial condition containing a monopole charge produces tails characterized by exponent $-3/2$, while an initial condition containing dipole charge produces tails characterized by exponent $-5/2$. In the presence of dipole conservation (our work) the exponents are inverted - an initial state with monopole charge produces tails characterized by an exponent $-5/2$, whereas an initial state with dipole charge produces tails characterized by exponent $-3/2$.}

\subsection{Entanglement diagnostics}

A question closely related to operator hydrodynamics is the problem of entanglement growth.  An initial product state develops spatial entanglement as it evolves with time. In systems without quenched disorder, the entanglement entropy is expected to grow linearly in time, with a growth rate characterized by the ``entanglement velocity'', $v_{E}$.  This entanglement growth generically continues until the system has reached a maximally entangled thermal state.  We here investigate entanglement growth in the fractonic circuit using three different entanglement metrics - the observable entanglement entropy, the operator entanglement, and the entanglement spectrum. We note that the conventional von Neumann entropy is not a suitable measure, because it is identically zero for a pure state, and nor is the conventional entanglement entropy (von Neumann entropy of a bipartition), because according to this measure the initial condition we consider is close to maximal entanglement. 
 
\subsubsection{Growth of local observable entanglement entropy}
\begin{figure}[b]
    \centering
    \includegraphics[scale=0.38]{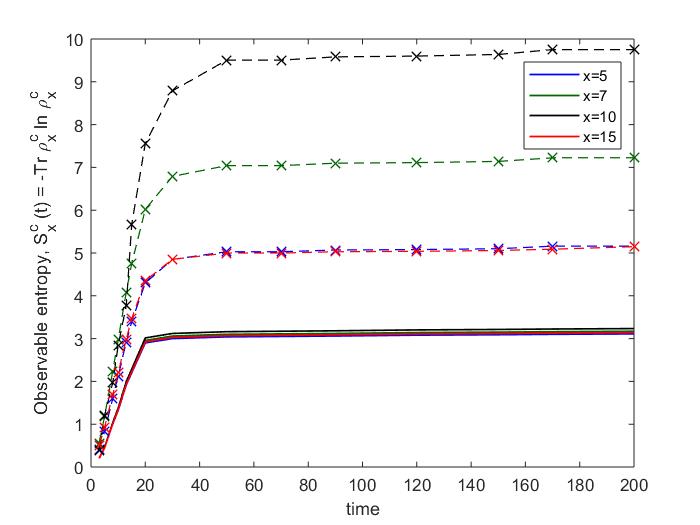}
  \caption{Local ``observable'' entanglement entropy, $S^{c}_{x}(t)$ (averaged over $10$ runs) as a function of time in the spin-1 system ($L=18$) with dipole conservation (solid lines) and without (dashed lines).}
   \label{fig:ententropy}
\end{figure}

We first work with ``observable" entanglement entropy\cite{KhemaniVishHuse}, arising from the conserved piece of the initially localized operator.  We compute the difference between the typical value of the local observable  entropy between two sides of a spatial entanglement cut in our system at time $t$ and the value of local observable entanglement entropy at $t=0$, and extract the entanglement velocity from it by looking at how the growth of observable entropy of the spin-$1$ system scales with time. To define the observable entropy more explicitly, consider the density matrix $\rho(t)= (\mathbb{I} + O(t))/3^{L}$. The von Neumann entropy of this state is independent of time by unitarity. Now, if we consider the conserved part of this state, i.e. $\rho^{c}(t)= (\mathbb{I} + O^{c}(t))/3^{L}$, then the observable entropy of this state is $S^{c}(t) = -\textrm{Tr}[\rho^{c}(t)\textrm{log}\rho^{c}(t)]$. We note that while the growth of local observable entanglement entropy scales linearly with $t$ at intermediate times, it saturates at large $t$. We determine the growth of the local observable entanglement entropy as a function of time for four cuts at location $x$ with $x=5, 7, 10, 15$ in an $L=21$ spin-1 chain (Figure \ref{fig:ententropy}), and see that while the initial scaling of the growth of local observable entanglement entropy is the same in all three cases, the entanglement velocity ($ \sim 0.16$) does not match the front velocity ($ \sim 0.4$) (Figure \ref{fig:frontposition}). This is normal and well understood - see e.g. Ref.\cite{Keyserlingk2}. 

Our numerical results also show a distinct difference in the late-time growth of local observable entanglement entropy of the fractonic system versus a system of ordinary conserved charges.  For ordinary charge conservation, the final saturation value of the growth of local observable entanglement entropy scales linearly with the size of the partition, indicating a volume law for growth of entanglement, as expected for a thermalizing state.  Specifically, the saturation value is very close to maximal observable entanglement.  In contrast, the growth of local observable entanglement entropy in a fracton system stops well short of its maximal value, and the saturation value is largely independent of partition size, indicating an area law for the growth of entanglement.  Such an area law for the growth of local observable entanglement entropy is consistent with the fact that a low-dimensional fracton system fails to thermalize under random unitary dynamics.

\subsubsection{Athermal operator entanglement}
We now investigate a different metric for the production of operator entanglement within the framework of spreading operators in the fractonic circuit, which does not separate out the conserved piece of operators. It had been suggested that, according to such a metric, a generic operator rapidly becomes maximally entangled within the region where it is present \citep{HoAbanin}, but we find that this is not the case for fracton operators: both the dipole and fracton charge spreading operators are volume-law entangled, but the late-time entanglement entropy of the fracton charge operator is \emph{well below} that of a maximally entangled operator.

\begin{figure}[b]
    \centering
    \includegraphics[scale=0.55]{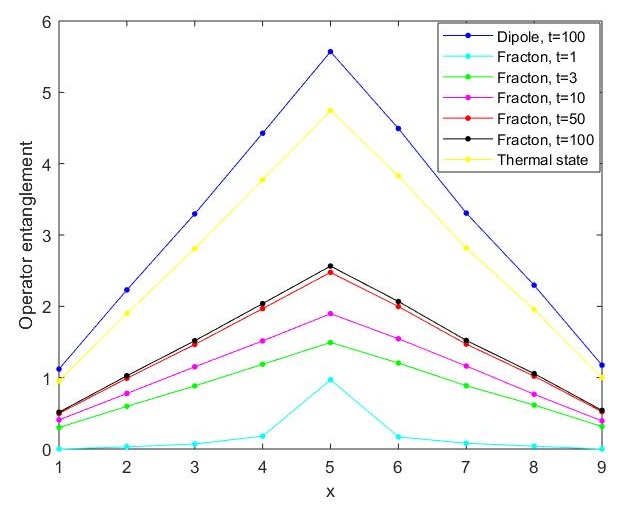}
  \caption{Operator entanglement (averaged over $10$ runs) for the fracton charge operator in a spin chain with $L=9$. The late-time operator entanglement of the corresponding thermal state and for the state evolved without the dipole constraint are also plotted, and these are well above the corresponding value for the fracton charge operator. The ``thermal'' state here is the equal weight superposition over all $S^z$ product states with a given charge and dipole moment. }
   \label{fig:opent}
\end{figure}

Our system has $3$ states per site: $|S_{z} = 0, \pm 1\rangle$.  Following the discussion in \citep{jonay2018coarse}, we may view an operator as a state in a doubled system with $9$ states per site. In order to see this mapping between operators and states, we consider a single site operator $O_{ab} |a\rangle \langle b|$, with $a, b$ labeling the three states at each site. The corresponding state is $||O\rangle \rangle = O_{ab} |a\rangle \otimes |b\rangle$. The notation $|| . . .\rangle \rangle$ indicates that this state lives in the doubled system. We can now construct two complete basis
sets $\{\hat{A}_{i}\}$ and $\{\hat{B}_{i}\}$ which are orthonormal, and consist only of operators with a support on the subsystems $A$ and $B$ respectively (taken to be the two halves of the spin chain in this case). For any linear operator $O$ in the full tensor product Hilbert space of the two operator spaces spanned by $\{\hat{A}_{i}\}$ and $\{\hat{B}_{i}\}$, we have a unique decomposition for $O$ as follows:
\begin{equation}
O=\sum_{i,j} O_{ij} |\hat{A}_{i}\rangle \otimes |\hat{B}_{j}\rangle ,
\end{equation}
where $O_{ij} = (\hat{A}_{i} \otimes \hat{B}_{j},O)$ is given by the inner product on the operator space. 

For a unitary time evolution operator $U(t)$ (like in each time step of our random unitary circuit), we have $U_{ij} = (\hat{A}_{i} \otimes \hat{B}_{j},U(t))$. From this, we construct the operator reduced density matrix
\begin{equation}
\rho^{A,\textrm{op}}_{ij} = \sum_{k} U_{ik}(U^{\dagger})_{kj}
\end{equation}
We use this to define the operator entanglement entropy \citep{zhouLuitz} $S^{\textrm{op}} = -\textrm{Tr}(\rho^{A,\textrm{op}} \textrm{ln} \rho^{A,\textrm{op}})$.

\begin{figure}[t]
    \centering
    \includegraphics[scale=0.35]{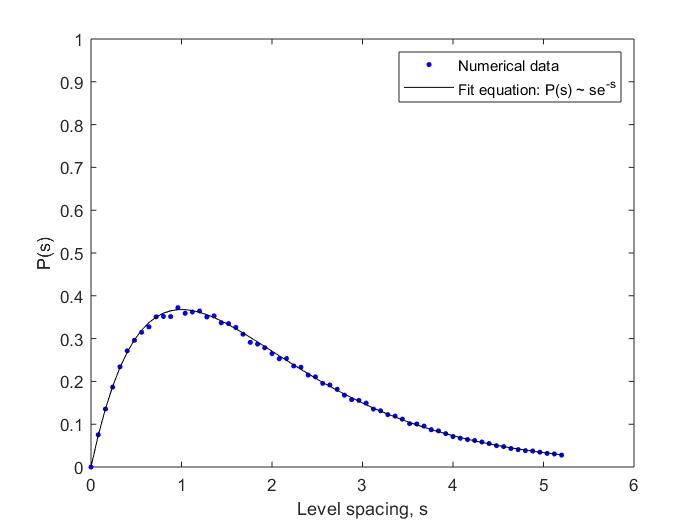}
  \caption{The level spacings of the entanglement Hamiltonian for the late-time state obey semi-Poisson statistics, in contrast with an integrable Anderson-like phase, which would obey Poisson statistics.}
   \label{fig:level}
\end{figure}
 
 
We note that while the operator entanglement for the fracton charge operator shows volume-law growth as a function of partition size, its late-time value is demonstrably lesser (see Figure \ref{fig:opent}) than the operator entanglement of the thermal state (a flat distribution over all $S^z$ product states with given charge and dipole moment).  This entanglement metric thereby serves as yet another independent confirmation that a system initialized with a single fracton fails to reach thermal equilibrium.

We note that the volume-law operator entanglement of the late-time configuration is in sharp contrast with the case of Anderson localization (or many-body localization), for which operator entanglement will remain in an area law at arbitrarily late times.  Furthermore, conventional localized phases would not have a ballistically propagating front, since there is no analogue of the emission of nonconserved operators.  Rather, conventional localized phases have a \emph{logarithmically} growing light cone \cite{mblarcmp}, resulting in a much slower spread of quantum information.

\subsubsection{Entanglement spectrum}
Another important question within a localized phase is the level statistics of the entanglement spectrum, $i.e.$ of the entanglement Hamiltonian $\tilde{H}$ defined by $\rho = e^{-\tilde{H}}$, where $\rho$ is the reduced density matrix for part of the system. We explore the entanglement spectrum by physically bipartitioning our system into a ``left'' and ``right'' half, constructing the reduced density matrix for the left half, and then obtaining the spectrum of the entanglement Hamiltonian. In Figure \ref{fig:level}, we plot the level spacings of the late-time state for a configuration initialized with a single fracton.  In a thermal spectrum, these level spacings would be expected to follow random matrix theory, whereas in a conventional integrable system they would be expected to be Poisson. We find a behavior that is neither random matrix nor Poisson, but which rather follows the {\it semi-Poisson} distribution, just like in many body localized systems \cite{es}.   The entanglement spectrum level statistics provide further evidence that this system is not thermalized, and nor is it conventionally integrable.

\subsection{Non-zero fracton density}

\begin{figure*}[t]
  \begin{subfigure}{8cm}
    \centering\includegraphics[scale=0.33]{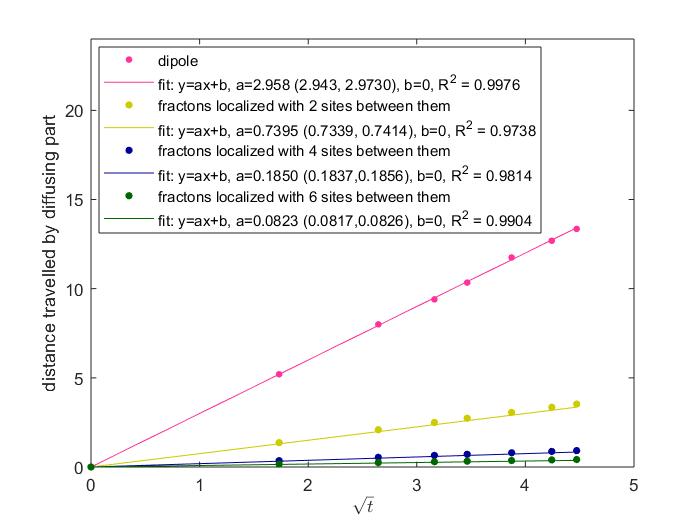}
    \caption{Diffusion constant of fractons in a two-fracton system is lower than that of the dipoles by a factor of $1/l^{2}$, where $l$ is the initial separation between the fractons. The uncertainties in the fitting parameters are given in the legend.}
  \end{subfigure}
    \begin{subfigure}{8cm}
    \centering\includegraphics[scale=0.33]{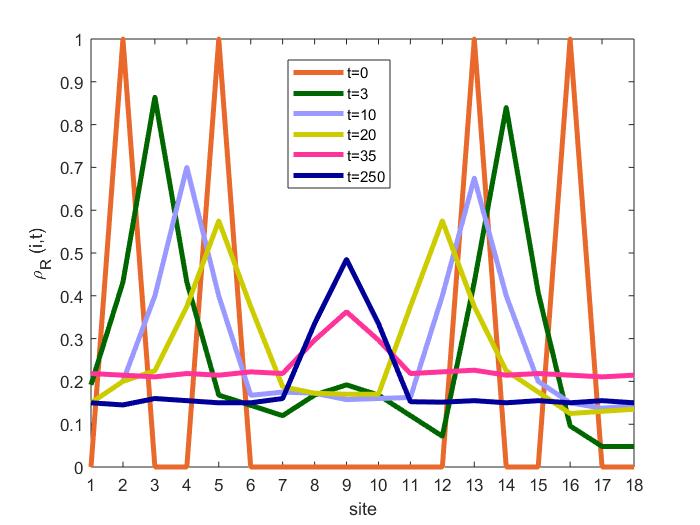}
    \caption{Right-weight profile in a system with a finite density of fractons (we look at the superposition of the following strings: $\mathbb{I}$ everywhere except at site $2$, $\mathbb{I}$ everywhere except at site $5$, $\mathbb{I}$ everywhere except at site $13$, $\mathbb{I}$ everywhere except at site $16$). Note that the fractons end up agglomerating at their center of mass.  }
  \end{subfigure}
  \caption{Fracton diffusion in the random circuit system initialized with multiple fracton peaks (averaged over $10$ runs).}
   \label{fig:finite}
\end{figure*}

We now turn to configurations having multiple fracton charge operators localized at different sites in the system. Now dipoles emitted during fracton motion do not have to return: they can be absorbed by the other fracton instead. In consequence, the fracton operators {\it can} now permanently change their position. For a system with multiple fractons, the diffusion constant for the fractons could be estimated as follows: for fractons a distance $l$ apart to move, they need to exchange dipoles. The diffusion time for dipoles is $~ D l^{2}$. Thus the fractons move an amount of the order $1$ in a time of order $Dl^{2}$. Thus, the diffusion constant of the fractons would be lower than the diffusion constant of the dipoles by a factor of $l^{2}$, where $l$ is the initial separation between fractons. This identification is supported by the numerical results in Figure \ref{fig:finite}a.

This basic picture has an interesting corollary: the closer together the fractons are, the more mobile they will be. The random circuit should thus generate an effective ``attraction'' between fractons (analogous to the discussion in \cite{mach}). In the long time limit, therefore, the fractons should agglomerate at their center of mass. This is confirmed by Figure \ref{fig:finite}b.  We further note an interesting ``coarsening'' dynamics to the system, in which nearby fracton peaks first coalesce together, before all peaks eventually join together at the center of mass of the system.  We expect such a coarsening process to lead to an infinite hierarchy of timescales for relaxation, for a finite density system in the thermodynamic limit. We also note that the state being relaxed to is one where the fractons all agglomerate near their center of charge, and this is very different from the classical thermal state (i.e. an equal weight superposition of all $S^z$ product states with a given charge and dipole moment, see Figure \ref{fig:thermalstate}). Thus, the system does not relax to an ergodic state, at least according to the intuitive definition of ergodicity. 

\begin{table*}[t]
  \centering
\begin{tabular}{ |P{4cm}|P{4cm}|P{4cm}|P{4cm}|  }
 \hline
  & Center of mass agglomerate & Thermal state in the charge $2$ sector & Two well-separated fracton peaks\\
 \hline
 Late-time operator entanglement   &  6.53   & 7.48 & 5.12  \\
 \hline
 \end{tabular}
\caption{Comparison of the late-time operator entanglement value of the agglomerate state at finite fracton density with those of the corresponding thermal state and of two well-separated fractons.}
  \label{opentcluster}
\end{table*}

The agglomeration of fractons at their center of charge also resolves a conceptual puzzle that may have worried the skeptical reader. Namely, random circuit dynamics allows for the creation from vacuum of dipole-anti-dipole pairs, and their subsequent dissociation into fractonic charges. Why then cannot fractons move by exchanging dipoles with such ``spontaneously generated'' fractons? The resolution is simple: even while fractonic charges can be created from vacuum, these creation processes cannot affect the total fracton charge or dipole moment, and hence cannot affect the center of charge. Since fracton charge ends up agglomerating at the center of charge, and the center of charge is a conserved quantity, such spontaneous pair production cannot affect the long time steady states of the system. 

\begin{figure}[b]
 \centering
 \includegraphics[scale=0.55]{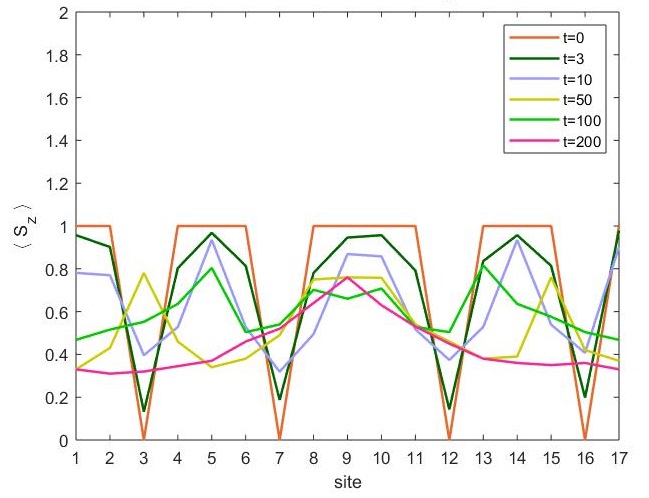}
 \caption{$\langle S_{z}(t)\rangle$ close to the maximal charge sector. The late-time profile shows a peak indicating localization even at finite fracton density.}
 \label{fig:maximalchargesector}
 \end{figure}
 
 \begin{figure}[b]
 \centering
 \includegraphics[scale=0.31]{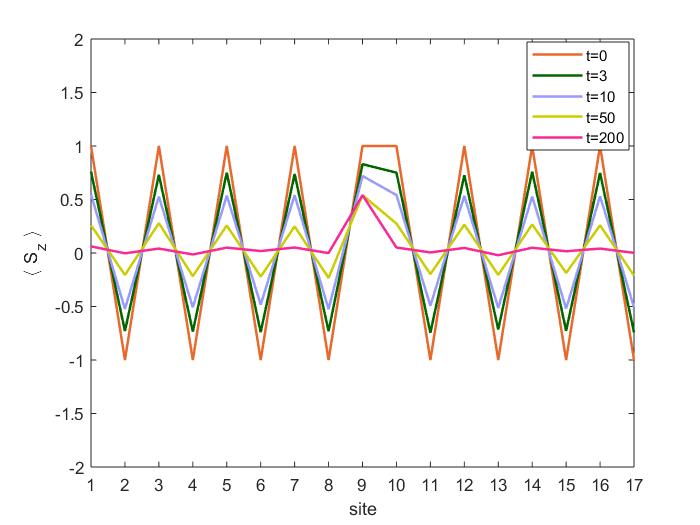}
 \caption{$\langle S_{z}(t)\rangle$ for an initial condition in the single charge sector but packed with dipoles. The late-time profile shows a peak indicating localization even in this limit.}
 \label{fig:fullDipole}
 \end{figure}
  
We now propose a possible mechanism for the interfracton attraction. While the analogy to the ``gravitational'' attraction between fractons discussed in \cite{mach} is appealing, a precise connection is difficult, since the discussion in \cite{mach} relies on the existence of a conserved energy. Instead, we propose that this attraction may be of entropic origin, and may be driven by the spreading of the non-conserved part of the operator. The appeal to an entropic mechanism may be surprising, since of course the {\it charge} sector does {\it not} thermalize, but the ``non-conserved'' parts of the operator {\it do} spread, and presumably tend towards their maximum entropy configuration for a given ``charge'' profile, and we propose that having the charge agglomerate at its center of mass increases the entropy of the ``non-conserved'' part of the operator, even though it does {\it not} correspond to the maximum entropy in the charge sector \footnote{This suggests that the ``fracton agglomeration'' is a consequence of feedback from the non-conserved sector onto the conserved sector. Curiously, this feedback is suppressed in the large $q$ limit discussed in \cite{KhemaniVishHuse}, suggesting that the fracton agglomeration phenomenon may not be accessible through analytical techniques like those from from \cite{KhemaniVishHuse}}. 
  
If we take the total weight (of late-time operator right weights) held in two well-separated fracton peaks ($\approx 1$, see Figure \ref{fig:chargeRprofile}) and consider the situation where the two fractons have come together (Figure \ref{fig:finite}b), and look at the total weight in the operator right weight peak ($\approx 0.4$), we see that the total weight in the resulting peak is reduced when the two fractons come together. This suggests that fractons coming together allows the system to ``emit'' more right weight and thus have higher entropy in the ``non-conserved'' part of the operator. To further support our claim that clustering at finite fracton density occurs for entropic reasons, we compare late-time operator entanglement values for the different states in question in Table \ref{opentcluster}, and we see that the state where the fractons agglomerate at their center of mass has higher operator entanglement than twice the operator entanglement of a single fracton, but less than that of the corresponding thermal state (consistent with lack of thermalization in the charge sector). A complete understanding of the inter-fracton attraction and its consequences, however, is beyond this scope of this work. 

Finally, we have numerically verified that the fracton localization persists at finite fracton density. In Figure \ref{fig:maximalchargesector} we plot the evolution of $\langle S^z(r) \rangle$ starting from an initial condition that is close to maximal charge (i.e. packed full of fractons). We still observe localization of fracton charge at late times, despite the fact that this initial condition is most certainly at non-zero fracton density. Similarly, in Figure \ref{fig:fullDipole} we consider an initial condition that is packed full of dipoles, and again, we observe localization of fractonic charge at long times. We thus conclude that the localization phenomenon we have discovered here  persists to systems at non-zero density, with the long time limit of such systems being a state where charge is localized near the center of charge. 

\section{Discussion and Conclusions}
\begin{figure*}[t]
\centering
\begin{subfigure}{0.99\columnwidth}
\includegraphics[width=\columnwidth]{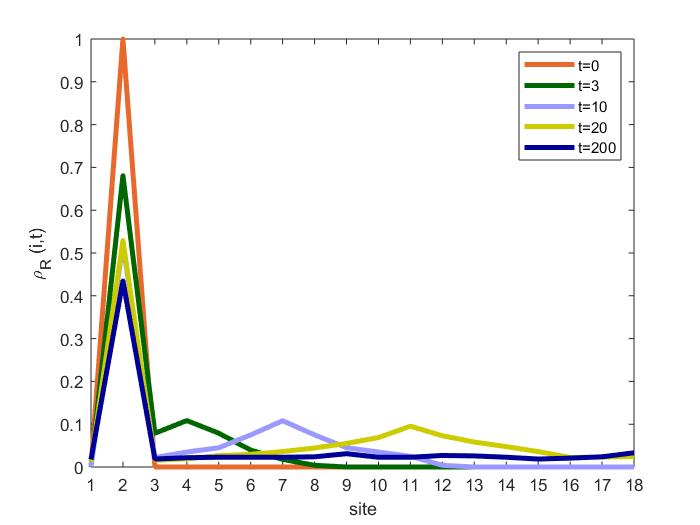}%
\caption{$L=18$ spin-$10$ chain}%
\end{subfigure}\hfill%
\begin{subfigure}{1.0\columnwidth}
\includegraphics[width=\columnwidth]{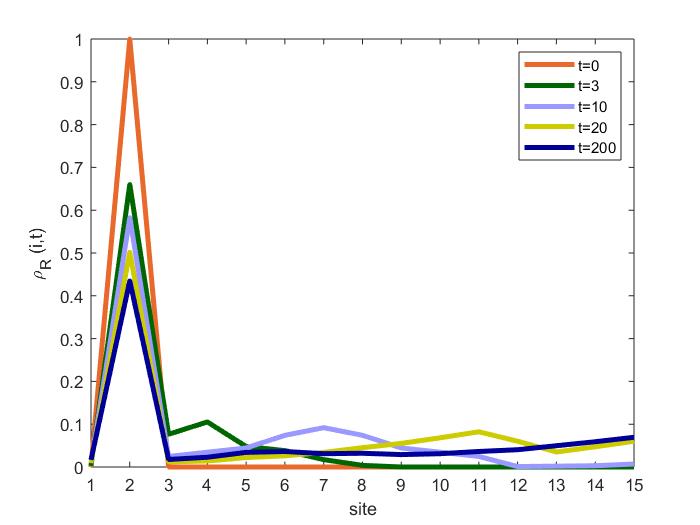}%
\caption{$L=15$ spin-$100$ chain}%
\end{subfigure}\hfill%
\caption{Right-weight profile (for a typical run) of the spreading fracton charge operator in a large $S$ system with open boundary conditions.}
\label{fig:largeSpeak}
\end{figure*}

We investigated the operator spreading dynamics of a one-dimensional random unitary circuit with fracton conservation laws, wherein we had not only a conserved $U(1)$ charge, but also the conservation of local dipole moment. We find that even though the circuit contains gates that {\it can} move fractonic charge, fractonic charge nevertheless does not move under random circuit dynamics. This remarkable ``localization'' of fractonic charge under random circuit evolution can be traced to the inevitable returns of low dimensional random walks.  In this sense, the mechanism is akin to weak localization, except that it is not based on quantum interference and is therefore insensitive to dephasing.  There is also some family resemblance to the diffusive returns governing the Altshuler-Aronov correction to conductivity in weakly-disordered metals \cite{aa}.  We note, however, that our system has neither a conserved energy nor any meaningful notion of density of states, so there is no direct parallel to e.g. \cite{aa, glass}. We have proposed a set of coupled hydrodynamic equations to describe the dynamics of our problem. Solution of these hydrodynamic equations predicts localization of charge in one and two spatial dimensions, but not in higher dimensions. It also makes a number of nontrivial predictions regarding the steady state distribution of charge for various initial conditions, and its dependence on the range of the gates. These predictions are in excellent agreement with numerical simulations in $d=1$, where high quality numerics is available, and we therefore have high confidence in our predictions for higher dimensions, where direct  simulation of the quantum dynamics is infeasible.


Even while fractonic charge fails to spread under random circuit dynamics, the non-spreading fractonic charge emits a ballistically propagating front of ``non-conserved'' operators, similar to the case of non-fractonic circuits \cite{KhemaniVishHuse}. The ballistically propagating front broadens as $\sqrt{t}$, again much as in non-fractonic circuits. However, the tail behind the front is characterized by a different power law to the case of non-fractonic circuits, and hence falls in a different universality class. These non-standard exponents are also correctly predicted by our hydrodynamic equations, and can be understood as a consequence of the unusual hydrodynamics governing the fractonic circuit, in which conserved ``dipolar'' charges are coupled to ``fractonic'' charges that acts as sources and sinks for dipole charge. 

We have considered the entanglement spreading in fractonic circuits. Since the initial conditions we consider are close to maximal entanglement according to the conventional entanglement entropy measure, we have to use different entanglement measures. We have considered three different entanglement measures: observable entanglement entropy, operator entanglement, and entanglement spectrum. By looking at observable entanglement entropy, we have found that while observable entanglement grows ballistically, the entanglement velocity is much smaller than the front propagation velocity. Additionally, a system initialized with fracton charge appears to saturate to an ``area law'' observable entanglement entropy independent of system size, consistent with the lack of thermalization. Meanwhile, the operator entanglement is volume law but subthermal, whereas the entanglement spectrum exhibits level statistics that follow a semi-Poisson distribution, similar to many body localized eigenstates \cite{es} but distinct to either thermalizing or conventionally integrable systems.  Finally, we have considered initial conditions that contain multiple fractons. If the initial condition contains multiple fractons, then the system evolves to a localized configuration where all fractons agglomerate at their ``center of charge'', and the rate of evolution to this configuration is determined by the initial separation of the fractonic charges. We have conjectured that this fracton agglomeration may be a consequence of feedback from the non-conserved operator to the conserved operator sector. Regardless of the origin of this attraction, however, the long time steady state attained is non-ergodic, even at non-zero fracton density $i.e.$ it is not equivalent to a thermal state (defined as an equal weight superposition of all $S^z$ product states with a given charge and dipole moment). Such a ``thermal'' state does not exhibit charge localization, and has markedly different entanglement properties. 
This constitutes the first example that we know of in which unitary time evolution with no time translation symmetry at all leads to localization.

We comment now on the implications of our results. Firstly, our work conclusively establishes that random circuit dynamics {\it can} lead to non-ergodic dynamical phases. This has obvious implications for quantum information processing. Particularly important is our prediction that localization in quantum circuits can persist to two dimensions, since actual information processing architectures cannot be purely one dimensional, such that our work opens up the possibility of harnessing localization for future scalable quantum circuits. Insofar as random circuit dynamics is a minimally structured model for time evolution, we expect the results to be robust, and applicable to any generic model with $U(1)$ charge conservation and dipole conservation, and no other constraints. 
\begin{figure}[b]
\centering
\includegraphics[width=\columnwidth]{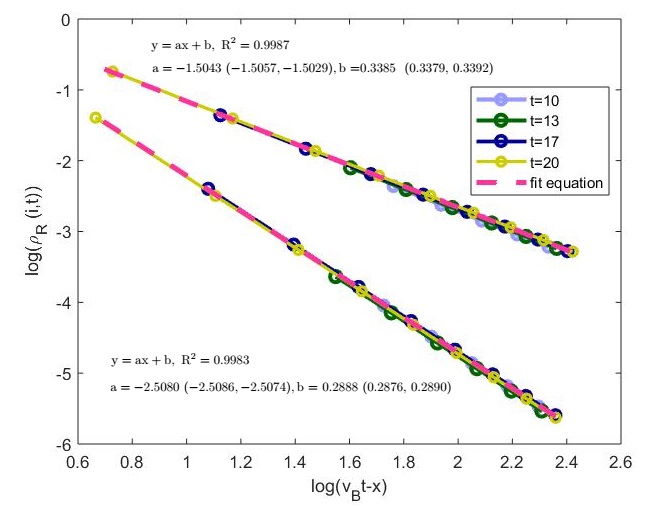}%
\caption{Power-law behavior of tails lagging behind propagating fronts (for a typical run) in a large $S$ ($S=10$, $L=18$) system with open boundary conditions. Uncertainties in fitting parameters are given in the plot.}
\label{fig:largeStail}
\end{figure}

Our work also has intriguing implications for the {\it Hamiltonian} dynamics of fracton systems. In particular, since Hamiltonian dynamics is more constrained than random circuit dynamics (because Hamiltonian dynamics must conserve energy), the immobility of fractons under random circuit dynamics in $d=1,2$ strongly suggests that fractonic systems in $d=1,2$ should also fail to thermalize under Hamiltonian dynamics. That is, fractonic systems in $d=1,2$, prepared in a sector with non-zero fracton charge but zero fracton charge density should have the fractons be strictly immobile, preserving forever a memory of their initial conditions in local observables - the very definition of MBL \cite{mblarcmp}. Moreover, this localization should survive to non-zero energy densities (unlike \cite{KimHaah}), since while initializing the system with a non-zero density of dipoles will ``break'' the locator expansion style arguments of \cite{KimHaah}, it will not affect the mechanism underlying the immobility of fractons in our analysis. The localization of fractonic charges will also be robust to arbitrary local perturbations - after all, it is robust to application of a random circuit. In other words, low dimensional fracton systems {\it can} realize MBL even in the absence of disorder in the Hamiltonian, and even in $d=2$. The obstructions to MBL in higher dimensions and in translation invariant Hamiltonians \cite{deroeck1, deroeck2} do not apply, since these are essentially obstructions to construction of a convergent locator expansion, and our argument for fracton localization does not rely on locator expansions. By thus liberating the study of MBL from reliance on the crutch of locator expansions, our work may also open a new chapter in this field. Moreover, the localization in low dimensional fractonic systems should be even more robust than traditional MBL, since the localization mechanism we have discovered herein is insensitive to noise, whereas conventional MBL does not survive noise \cite{NandkishoreGopalakrishnan}.  \footnote{The only fly in the ointment is that all known {\it Hamiltonian} $U(1)$ fracton models in lower dimensions exhibit electrostatic confinement \cite{sub}, so the information has to be encoded in infinitely energetic excitations. However, while we are not aware of any models that realize stable phases of $U(1)$ fractons in lower dimensions without electrostatic confinement (without engaging in fine tuning of the Hamiltonian), we are also not aware of any general arguments indicating that such models cannot exist. The search for suitable models in low dimensions would be an interesting topic for future work, with the present work serving as a direct motivation for such a search. }

\begin{figure}[t]
    \centering
    \includegraphics[scale=0.35]{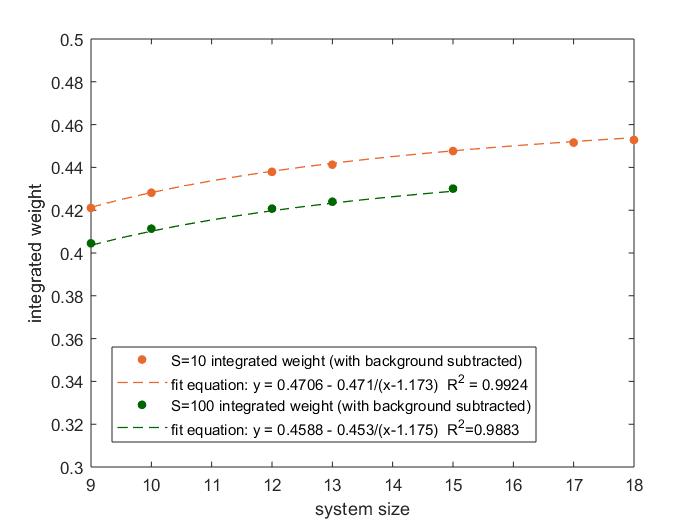}
  \caption{Integrated weight in the fracton peak in the right-weight profile for large $S$ at late times ($t \sim 200$).}
   \label{fig:intwtlargeS}
\end{figure}

Our work also opens up new possibilities for noisy fracton dynamics in three spatial dimensions (where fracton physics is best understood). Previous explorations of fracton dynamics in three dimensions (e.g. \cite{prem, SivaYoshida}) have essentially demonstrated the lack of convergence of ``locator expansion'' type approaches to fracton dynamics at non-zero energy densities. This lack of convergence has been interpreted in terms of thermalization, and the diffusion constant inferred from the breakdown of the locator expansion. However, in the present work we have identified a new mechanism for ergodicity breaking, which does not rely on locator expansions. This re-opens the possibility of a non-ergodic long time limit for three dimensional fracton systems. Prima facie, our analysis does not apply to three dimensional systems, since our analysis relies on the return probabilities of low dimensional random walks, and in three dimensions random walks need not return to the origin. However, the ``dipolar'' excitations in fractonic models can frequently be ``subdimensional'' \cite{sub}, being restricted to move in a space of lower dimensionality. The localization of fractonic charge is tied to the return probability of dipoles, and the subdimensional character of dipoles in certain three dimensional fracton models may open the door to  non-ergodic long time states for three dimensional fractons. This too is an interesting topic for future exploration. 

Finally, it is worth critically examining the role of quantum mechanics in the localization of fractons.  Throughout, we have made heavy use of the machinery of quantum mechanics, such as quantum states and operators.  Indeed, our results have several intrinsically quantum-mechanical features, such as operator spreading dynamics and unusual entanglement behavior. We have also invoked feedback between the conserved and non-conserved operator sectors, which does not have an obvious  classical analogue. However, our primary argument for localization, in terms of the guaranteed return of low-dimensional random walks, may well hold at the classical level as well.  It is therefore plausible that there may be an analogous classical phase featuring localized charge dynamics.  This would open the door to a new type of purely classical localization phenomenon, in contrast with the quantum mechanical origin of standard MBL phases.  We leave this question as a topic for future investigation.

\section*{Acknowledgments}
The authors acknowledge useful discussions with Yang-Zhi Chou, David Huse, Vedika Khemani, Abhinav Prem, Jonathan Ruhman, and Sagar Vijay. We thank Paul Romatschke and Ted Nzuonkwelle for providing access to the Eridanus cluster at CU Boulder. This material is based upon work supported by the Air Force Office of Scientific Research under award number FA9550-17-1-0183 (SP, MP and RMN); by the U.S. Department of Energy, Office of Science, Basic Energy Sciences (BES) under Award number DE-SC0014415 (SP); a Simons Investigator Award to Leo Radzihovsky (MP); by NSF Grant 1734006 (MP), and by the Foundational Questions Institute (fqxi.org; grant no. FQXi-RFP-1617) through their fund at the Silicon Valley Community Foundation (RN and MP).  We also acknowledge the support of the Alfred P. Sloan foundation through a Sloan Research Fellowship (RN).

\section*{Appendix: Large $S$ limit}
Here we look at what happens to the non-ergodic phenomenon discussed in the main text for a spin-$S$ ($S \neq 1$) chain of length $L$ in one dimension. To extend the above model to large $S$, we use generalized Gell-Mann matrices of dimension $2S+1$, i.e. $\bigg(\Sigma_{i}^{1,...,(2S+1)^{2}-1}\bigg)$ along with the $(2S+1) \times (2S+1)$ identity matrix $\bigg(\Sigma_{i}^{0}\bigg)$ as the $(2S+1)^{2}$ elements of the onsite operator basis. The onsite basis has the following normalization: $\textrm{Tr}\big[ \frac{\Sigma_{i}^{\mu \dagger}\Sigma_{i}^{\nu}}{2S+1}\big]=\delta_{\mu \nu}$. We use these matrices to form a basis of $(2S+1)^{2L}$ generalized Pauli strings $\tilde{S}=\prod_{i}\Sigma_{i}^{\mu_{i}}$. The initially-local operator $O_{o}$ consists only of those strings that are the identity at all sites except a few near the site of initialization. With time, however, there are other basis strings with nonlocal weight in the decomposition of the time-evolved operator $O(t)$.

We include numerical results for $S=10$ and $S=100$ spin chains subject to random unitary evolution via $3$-qudit gates. Figure \ref{fig:largeSpeak} shows the fracton peak in the right-weight profiles for $S=10$ and $S=100$. In Figure \ref{fig:largeStail}, we show the power-law behavior of the tails lagging behind the front and provide numerical evidence for the $-5/2$ exponent appearing even at large $S$. And finally, we plot the integrated weight under the fracton peak as a function of system size, and see that it saturates to a finite value in the thermodynamic limit in both cases (see Figure \ref{fig:intwtlargeS}). Altogether we conclude that our basic results survive as we take the limit of large $S$. 

\bibliography{library}

\newpage

\section{Erratum to localization in fractonic random circuits}

We report some errors in the published version of our paper \cite{prx}. 

\begin{enumerate}

\item There is a typo in the bottom right box of Table 1. Specifically, the state (+-+-) should have transitions with (+0-0) and (0+0-), and not with (+00-). The table should also include transitions between (-0+0), (0-0+), and (-+-+). These typos in the table were not reflected in our actual simulations, which featured all correct gates.

\item The caption to Figure 21 incorrectly states the initial condition to be a mixed state, of the form ($I\otimes S_z\otimes I\otimes I\otimes I...+ I\otimes I\otimes I\otimes S_z\otimes I...$) etc. Rather, the initial conditions for all simulations in this section are pure states, of the form $|0+00+0\rangle$ etc. This applies to the simulations depicted in Figures 14, 21, 22, and 23. Similarly, at the start of Section III.A, the initial condition for Fig.2 is incorrectly given to be of the
form $I\otimes S_z\otimes I\otimes I\otimes I...- I\otimes I\otimes S_z\otimes I \otimes I...$. Instead, the initial condition used was a pure state of the form $|0+-000...\rangle$. When considering pure states, the right weight was simply defined as the probability that the state was non-zero on that site. However, all simulations involving a single fracton (i.e. most of the paper) did indeed consider an initial condition of the form $I\otimes S_z \otimes I \otimes I ...$, and defined right weight as in \cite{Nahum2, Keyserlingk2}. 

\item The original version of this article stated that SP was supported by the U.S. Department of Energy, Office of Science, Basic Energy Sciences (BES) under Award number de-sc0014415 and
by the Air Force Office of Scientific Research under award number FA9550-17-1-0183. This was in error. For this work, SP was supported only by the Air Force Office of Scientific Research under award number FA9550-17-1-0183.

\item Finally, subsequent to the publication of this work, we became aware that when our simulations were run on the cluster, they were subject to an uncontrolled truncation when certain basis string coefficients became smaller than machine precision. This is evidenced by the fact that the sum of squares of basis string coefficients is less than one. (e.g. for $L=21$ and four site gates it is $0.98$). In addition, there can arise a truncation when the  memory requirements exceed the available memory.  This latter truncation does not show up in the sum of squares of string coefficients, but may also have affected our results. To determine whether our results were qualitatively affected by truncation, we have double checked our results using an independently written `brute force' code run on small enough system sizes that such truncation is not an issue.  We find that our key results for the core model of spin $S=1$ and three site gates, which constitute the vast majority of the paper, are largely unchanged, but truncation issues may have affected our conclusions regarding robustness of localization in models with longer range gates and/or large spins.  We have also triple checked our results using an independent matrix product state (MPS) code supplied by our colleague Dr. Jason Iaconis. Our conclusions are below. 

\begin{figure*}[t]
  \begin{subfigure}{5.5cm}
\centering
 \includegraphics[scale=0.35]{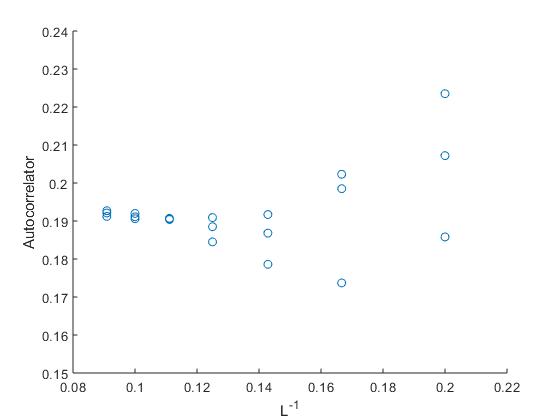}
 \caption{Autocorrelator as a function of inverse system size for three site gates. }
 \label{fig:rightweight}
 \end{subfigure}
 \hspace{3cm}
 \begin{subfigure}{5.5cm}
\centering
 \includegraphics[scale=0.35]{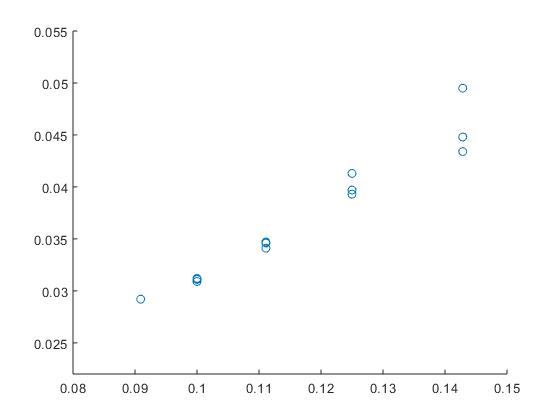}
 \caption{Autocorrelator as a function of inverse system size for four site gates.}
 \label{fig:SvsEall}
 \end{subfigure}
 \caption{Late-time value of the autocorrelation function, $\langle S_{z,i}(t)S_{z,i}(0)\rangle - \frac{2}{3L}$, as a function of inverse system size, $L^{-1}$, for 3-site and 4-site gates.  Results obtained with the brute force code, with $t=1000$. The correlation function is defined as $\langle S_{z,i}(t)S_{z,i}(0)\rangle = \frac{1}{3^L} Tr(S^z_i O(t))$, where $O(0) = ...I \otimes I \otimes S^z \otimes I \otimes I...$, and the initial $S^z$ is on a site $i$ in the middle of the chain. In both cases, three runs are plotted for each system size to illustrate the size of fluctuations, which become negligible at larger system sizes.}
 \end{figure*}

For the minimal model of a fractonic circuit with three site gates, which is the model considered in most of our paper, our key results are essentially unchanged. In particular, we have independently verified our signature result, that initial conditions of the form $I\otimes S_z\otimes I\otimes I\otimes I$ lead to localization of charge, up to system size $L=11$ (see Fig.1a). Our original simulations for operator right weight also observed anomalous exponents of $3/2$ and $5/2$ governing the decay of right weight in the central peak, and the `tails behind the propagating front' respectively. We have verified using the MPS code for system size $L=9$ that the decay of the central peak is governed by the anomalous exponent of $3/2$. While the system sizes that we can access without truncation are not large enough for a respectable fit to the `tail behind the front,' the two exponents are tied together \cite{KhemaniVishHuse}, so that an anomalous exponent $3/2$ for the central peak implies an exponent $5/2$ for the tail behind the front.  Finally, we have verified fracton agglomeration by examining $\langle S^z(r)\rangle$ as a function of time for pure state initial conditions containing multiple fractons up to $L=9$.   As such there are no significant changes to our analysis of the model with three site gates, which constitutes the bulk of our paper.
There are some minor differences which may be due to truncation (e.g. the right weight distribution for the small sizes we can access has more of an upturn towards the end of the chain than in Fig.3 of \cite{prx}), but the essential results are unaffected.

For fractonic circuits with longer ranged gates, our paper originally claimed localization survived, with a characteristic dependence of localization length on system size (Fig.5, 13). However, \cite{munich} suggested that this may have been purely a finite size effect. Motivated by this tension between our results and those of \cite{munich}, we have reexamined our results on localization with longer range gates, using both independently written codes, on system sizes small enough for truncation to not be an issue. Specifically, we have plotted the quantity suggested by \cite{munich}, $\langle S^z_i(0) S^z_i(t) \rangle - 2/3L$, for an initial condition of the form $...I \otimes I\otimes S_z\otimes I\otimes I\otimes I$ in a chain of size $L$, with the $S^z$ on a site $i$ chosen in the {\it middle} of the chain. The results are shown in Fig.1b. While numerical extrapolation to the thermodynamic limit may give something non-zero, it nevertheless gives a value that is exceedingly small i.e. {\it most} of the charge escapes. Consequently, it is not clear that it makes sense to talk about localization lengths for longer range gates. This also suggests that the hydrodynamic picture (Eq.8) cannot be complete. What terms need to be added to the equations to capture the effect of longer range gates is an interesting open question. 

Finally, in the Appendix we presented results on right weights for systems with large spin $S=10, 100$. We have not been able to reproduce these without truncation for system sizes large enough to say anything meaningful. Accordingly, we cannot exclude the possibility that these large $S$ results may be artifacts of truncation. 
\end{enumerate}

{\bf Acknowledgements} We thank Frank Pollmann, Vedika Khemani, and Tibor Rakovszky for alerting us to the possibility of truncation in our numerics. We are especially grateful to our colleague Jason Iaconis for his assistance in double checking our numerics.

\end{document}